\documentclass[11pt]{article}
\usepackage[fleqn]{amsmath}
\usepackage{amsthm,amssymb}
\usepackage{a4}
\usepackage{cite}
\makeatletter
\@addtoreset{equation}{section}
\makeatother

\newcommand{\be}[1]{\begin{equation}\label{#1}}
\newcommand{\ba}[1]{\begin{eqnarray}\label{#1}}
\newcommand{\bal}{\begin{align}}
\newcommand{\bmul}[1]{\begin{multline}\label{#1}}

\newcommand{\ee}{\end{equation}}
\newcommand{\ea}{\end{eqnarray}}
\newcommand{\eal}{\end{align}}
\newcommand{\emul}{\end{multline}} 

\newcommand{\eq}[1]{(\ref{#1})}

\newtheorem{thm}{\sc Theorem}[section]
\newtheorem{prop}{\sc Proposition}[section]

\newtheorem{lemma}{\sc Lemma}[section]
{\theoremstyle{remark}
\newtheorem{rem}{Remark}[section]}
\newtheorem{Def}{\sc Definition}[section]

\newcommand{\bra}[1]{\langle\,#1\,|}
\newcommand{\ket}[1]{|\,#1\,\rangle}
\newcommand{\bracket}[1]{\langle\,#1\,\rangle}

\def\tr{\operatorname{tr}}

\def\Res{\operatorname{Res}}

\def\sh{\sinh}
\def\ch{\cosh}

\def\pour#1{_{\,\vrule height 13pt depth 1pt\> {#1}\!}}

\def\sul{\sum\limits}
\def\pl{\prod\limits}

\def\build#1_#2^#3{\mathrel{
\mathop{\kern 0pt#1}\limits_{#2}^{#3}}}

\newcommand{\Zset }{{\mathbb Z}}

\newcommand{\Cset }{{\mathbb C}}

\usepackage{euscript}

\def\sg{\sigma}
\def\la{\lambda}

\def\tend{\rightarrow}

\let\qed=\cqfd

\newcommand{\negspace}{\!\!\!\!}
\newcommand{\Negspace}{\negspace\negspace}

\begin{document}

\begin{titlepage}
\begin{flushright}
LPENSL-TH-04/04\\
\end{flushright}

\vspace{24pt}

\begin{center}
\begin{LARGE}
{\bf On the algebraic Bethe Ansatz approach
 to the correlation functions
 of the $XXZ$ spin-$1/2$ Heisenberg chain}
\end{LARGE}

\vspace{50pt}

\begin{large}
{\bf N.~Kitanine}\footnote[1]{LPTM, UMR 8089 du CNRS,
Universit\'e de Cergy-Pontoise,
France, kitanine@ptm.u-cergy.fr},~~
{\bf J.~M.~Maillet}\footnote[2]{ Laboratoire de Physique, UMR 5672 du CNRS,
ENS Lyon,  France,
 maillet@ens-lyon.fr},~~
{\bf N.~A.~Slavnov}\footnote[3]{ Steklov Mathematical Institute,
Moscow, Russia, nslavnov@mi.ras.ru},~~
{\bf V.~Terras}\footnote[4]{LPMT, UMR 5825 du CNRS,
Universit\'e de Montpellier II, France, terras@lpm.univ-montp2.fr} \par
\end{large}

\vspace{80pt}

\centerline{\bf Abstract} \vspace{1cm}
\parbox{12cm}{We present a review of the method we have elaborated
to compute the correlation functions of the $XXZ$ spin-$1/2$ Heisenberg chain.
This method is based on the resolution of the quantum inverse scattering
problem in the algebraic Bethe Ansatz framework, and leads to a multiple
integral representation of the dynamical correlation functions.
We describe in particular some recent advances concerning the two-point
functions: in the finite chain, they can be expressed in terms of a
single multiple integral.
Such a formula provides a direct analytic connection between the previously
obtained multiple integral representations and the form factor expansions for
the correlation functions.
}
\end{center}

\vfill \noindent {This review article is based on a series of four
lectures given at the workshop {\em Recent Progress in Solvable
lattice Models}, RIMS Sciences Project Research 2004 on {\em
Method of Algebraic Analysis in Integrable Systems}, RIMS, Kyoto,
July 2004.}
\end{titlepage}
\newpage


\section{Introduction}\label{sec-1}

Computing exact and manageable expressions for correlation
functions is a central question in the theory of quantum
integrable models \cite{Bax82L,Gau83L,LieM66L}. This problem is of
great importance, both from a theoretical and mathematical view
point and for applications to various interesting physical
situations. Apart from few cases, like free fermions
\cite{Ons44,LieSM61,McC68,WuMTB76,McCTW77a,SatMJ78m} or conformal
field theories \cite{BelPZ84}, this issue is still far from its
complete solution. Although several important advances have been
obtained in the recent years, we are still lacking a general
method that could give in particular a systematic way to evaluate
compact expressions for two-point functions and their long
distance asymptotic behaviour.

The aim of the present paper is to give a review of an approach to
this problem elaborated in \cite{KitMT99,KitMT00,MaiT00} and in
\cite{KitMST02a,KitMST02b,KitMST02c,KitMST02d}, together with an
account of the more recent progress obtained in
\cite{KitMST04a,KitMST04b,KitMST04c}.

In our search for a general method to compute correlation
functions of quantum integrable systems, our strategy is to
consider a simple but representative model for which it is
possible to develop new concepts and tools towards this goal. An
archetype of such a model is provided by the $XXZ$
spin-$\frac{1}{2}$ Heisenberg chain in a magnetic field with
Hamiltonian,
\be{IHamXXZ} H=H^{(0)}-hS_z, \ee
where
\begin{align}\label{H0}
 &H^{(0)}=\sum_{m=1}^{M}\left\{
    \sigma^x_{m}\sigma^x_{m+1}+\sigma^y_{m}\sigma^y_{m+1}
    +\Delta(\sigma^z_{m}\sigma^z_{m+1}-1)\right\},\\
 &S_z=\frac{1}{2}\sum_{m=1}^{M}\sigma^z_{m}, \qquad
                   [H^{(0)},S_z]=0.
   \label{Sz}
\end{align}
Here $\Delta$ is the anisotropy parameter, $h$ denotes the
external classical magnetic field, and $\sigma^{x,y,z}_{m}$ are
the local spin operators (in the spin-$\frac12$ representation)
associated with each site of the chain. The quantum space of
states is ${\cal H}={\otimes}_{m=1}^M {\cal H}_m$, where ${\cal
H}_m\sim \mathbb{C}^2$ is called local quantum space. The
operators $\sigma^{x,y,z}_{m}$ act as the corresponding Pauli
matrices in the space ${\cal H}_m$ and as the identity operator
elsewhere. For simplicity, the length of the chain $M$ is chosen
to be  even and we assume periodic boundary conditions. Since the
simultaneous reversal of all spins is equivalent to a change of
sign of the magnetic field,  it is enough to consider the case
$h\ge 0$. In the thermodynamic limit $M\to\infty$ and at zero
magnetic field, the model exhibits  different regimes depending on
the value of $\Delta$ \cite{Bax82L}. The ground state is
ferromagnetic for $\Delta <-1$, while it has magnetisation zero
for $\Delta>-1$. In the last case the spectrum is gapless for $-1
< \Delta < 1$ (massless regime), while for $\Delta > 1$ the ground
state is twice degenerated with a gap in the spectrum (massive
regime).

We are basically interested in the two-point correlation functions
of local spins, although the results presented here allow us to
consider other correlation functions as well. If we restrict
ourselves to the zero temperature situation, such a problem comes
down to the computation of the average value of a product of two
local spin operators in the ground state $\ket{\psi_g}$ of the
Hamiltonian~\eq{IHamXXZ}:
\be{Defzz}
  g_{\alpha\beta}(m)=\bra{\psi_g}\,\sigma^\alpha_{1}\,
        \sigma^\beta_{m+1}\,\ket{\psi_g},
  \qquad (\alpha,\beta)=(+,-),(-,+),(z,z).
\ee
Despite its apparent simplicity, such an object is highly
non-trivial to handle. The first problem to solve is obviously to
determine the ground state $\ket{\psi_g}$. A method to diagonalise
the Hamiltonian was proposed by Bethe in 1931 \cite{Bet31} and
developed later in \cite{Orb58,Wal59,YanY66,YanY66a}. The
algebraic version of the Bethe Ansatz was created in the framework
of the Quantum Inverse Scattering Method by L.D. Faddeev and his
school \cite{FadST79,FadT79,BogIK93L}. Different ways to study the
correlation functions  of this model were proposed in the series
of works (see e.g. \cite{JimMMN92,JimM96,JimM95L,KitMT99,
KitMT00,KitMST02a,KitMST02b,KitMST02c,KitMST02d,BooKS04,GohKS04,BooJMST04}).

Multiple integral representation for the correlation functions
were obtained for the first time from the $q$-vertex operator
approach (also using corner transfer matrix technique) in the
massive regime $\Delta \ge 1$ in 1992 \cite{JimMMN92} and
conjectured in 1996 \cite{JimM96} in the massless regime $-1 \le
\Delta \le 1$ (see also \cite{JimM95L}). A proof of these results,
together with their extension to non-zero magnetic field, was
obtained in 1999 \cite{KitMT99,KitMT00} for both regimes using
algebraic Bethe Ansatz and the actual resolution of the so-called
quantum inverse scattering problem \cite{KitMT99,MaiT00}. In fact,
these multiple integral representations have been constructed for
the elementary building blocks (see Section~\ref{sec-2}), since
any arbitrary correlation function can be expressed in terms of a
linear combination of such blocks. One should note however that,
although these formulas are quite explicit, the actual analytic
computation of the corresponding multiple integrals is missing up
to now. Moreover, the evaluation of the two-point correlation
functions \eq{Defzz} at lattice distance $m$ is a priori quite
involved, since the number of terms in the corresponding linear
combination of the elementary blocks growths exponentially with
$m$ (like $2^{m}$). This makes the problem of asymptotic behaviour
at large distance extremely difficult to solve in this settings
from the present knowledge of the elementary blocks.

In the articles \cite{KitMST02a,KitMST04c} we have derived new
multiple integral representations more adapted to the two-point
correlation functions. One of them \cite{KitMST04c} is based on
the direct re-summation of the above linear combinations of the
elementary blocks. The second one \cite{KitMST02a} uses an
explicit representation for the multiple action of the twisted
transfer matrices (see \eq{Twis-T-M}) on an arbitrary Bethe state.
In both cases the number of multiple integrals describing the
two-point functions \eq{Defzz} reduces from $2^m$ to $m$.

The development of these methods allowed us to perform further
re-summation and to obtain representations for the two-point
functions on the lattice in terms of a single multiple integral
\cite{KitMST04a}. We call this type of representation {\sl master
equation}. The remarkable property of the master equation is that
it gives a direct analytic link between two general approaches to
the computation of correlation functions: in the context of the
$XXZ$ Heisenberg chain, the first method consists in acting with
the local operators $\sigma_1^\alpha$ and $\sigma_{m+1}^\beta$ on
the ground state $\bra{\psi_g}$ to produce a new state
$\bra{\psi(\alpha,\beta,m)}$, and then in computing the resulting
scalar product $\bracket{\psi(\alpha,\beta,m)\mid\psi_g}$; the
second method consists in inserting between the two operators of
local spin a sum over a complete set of states (for example
eigenstates of the Hamiltonian), which gives a decomposition of
the two-point function in the form
\be{Defzztff}
  g_{\alpha\beta}(m)=\sum_i \bra{\psi_g}\,\sigma^\alpha_{1}\,\ket{i}
     \bra{i}\, \sigma^\beta_{m+1}\,\ket{\psi_g}.
\ee
Using the technique developed in \cite{KitMST04a,KitMST04b}, we
are able to re-sum completely the form factor expansion
\eq{Defzztff} and to show that it leads indeed to the master
equation obtained by the first method. In fact, these two
different approaches have a very simple interpretation in the
context of the master equation. Namely, they correspond to two
different ways to evaluate the contour integral, by computing the
residues in the poles that are either inside or outside the
integration contour. The first way leads to a representation of
the correlation function
$\bracket{\sigma_1^\alpha\,\sigma_{m+1}^\beta}$ in terms of the
previously obtained \cite{KitMST02a} multiple integrals. The
second one gives us the form factor type expansion of the
correlation function  (i.e. an  expansion in terms of matrix
elements of local spin operators between the ground state and all
excited states).

This method  was generalised in  \cite{KitMST04b} to the
time-dependent (dynamical) correlation functions
\be{Defzzt}
 g_{\alpha\beta}(m,t)
    =\bra{\psi_g}\,\sigma^\alpha_{1}(0)\,
                \sigma^\beta_{m+1}(t)\,\ket{\psi_g}
    =\bra{\psi_g}\,\sigma^\alpha_{1}\, e^{iHt}\,
            \sigma^\beta_{m+1}\, e^{-iHt}\,\ket{\psi_g}.
\ee
It is worth mentioning that, up to now, the only known exact
results on the dynamical correlations concern the case of free
fermions $\Delta=0$
\cite{LieSM61,Nie67,McC68,McCBA71,PerC77,SatMJ78m,McCPS83,ColIKT93}.
It turns out, however, that the methods developed in
\cite{KitMST02a,KitMST04a} can be directly applied to the
computation of the time-dependent correlation functions. In
particular, one can obtain a time-dependent analogue of the master
equation and also a multiple integral representation for
$g_{\alpha\beta}(m,t)$ in the thermodynamic limit, both in massive
and massless regime.

This paper is organised as follows. In Section~\ref{sec-2}, we
briefly recall how to obtain multiple integral representations for
the elementary building blocks of the correlation functions using
the algebraic Bethe Ansatz method \cite{KitMT99,KitMT00,MaiT00},
and introduce useful techniques and notations that will be used
all along the article. In Section~\ref{sec-3} we explain how to
re-sum these elementary building blocks to obtain compact
representations for the two-point functions and their generating
functions \cite{KitMST02a,KitMST02b}. The problem of asymptotic
behaviour for large distances is tackled in Section~\ref{sec-4}.
There we consider the toy example of the so-called emptiness
formation probability to show how the multiple integral
representations of Section~\ref{sec-3} can be analysed in the
asymptotic limit of large distances, both in  the massless
\cite{KitMST02c,KitMST02d} and massive regimes. We also discuss
how the methods we have developed could be extended to the case of
the two-point functions. Section~\ref{sec-5} and \ref{sec-6} are
devoted to the derivation of the master equation for the
correlation functions by the two equivalent approaches that have
been mentioned above. In the last section we present our
conclusions and perspectives.


\section{Algebraic Bethe Ansatz and elementary blocks}\label{sec-2}

Any $n$-point correlation function of the Heisenberg chain can be
reconstructed as a sum of elementary building blocks  defined in
the following way:
\be{genabcd00}
   F_m(\{\epsilon_j,\epsilon'_j\})
    =
      \bra{\psi_g}\prod\limits_{j=1}^m
                      E^{\epsilon'_j,\epsilon_j}_j\ket{\psi_g}.
\ee
Here $\ket{\psi_g}$ is the normalised ground state of the chain
and $E^{\epsilon'_j,\epsilon_j}_{j}$ denotes the elementary
operator acting on the quantum space ${\cal H}_j$ at site $j$ as
the $2 \times 2$ matrix of elements
$E^{\epsilon',\epsilon}_{lk}=\delta_{l,\epsilon'}\delta_{k,\epsilon}$.

A multiple integral representation for these building blocks was
obtained for the first time in \cite{JimMMN92,JimM96}. In this
section, we briefly recall how it can be derived in the framework
of algebraic Bethe Ansatz \cite{KitMT99,KitMT00}. In general, we
have to solve the following successive problems:

$(i)$ determination of the ground state $\bra{\psi_g}$,

$(ii)$  evaluation of the action the product of local operators on
this ground state,

$(iii)$ computation of the scalar product of the resulting state
with $\ket{\psi_g}$,

$(iv)$ thermodynamic limit.

The starting point of our method is to use in step $(i)$ the
description of the eigenstates obtained via algebraic Bethe Ansatz
\cite{FadST79,BogIK93L}. They are constructed in this framework in
terms of generalised creation and annihilation operators which are
themselves highly non-local. Acting with local operators on such
states in step $(ii)$ is therefore a priori a non-trivial problem.
One of the key-ingredient of our method, which enables us to
compute this action explicitly, is the solution of the so-called
quantum inverse scattering problem \cite{KitMT99,MaiT00}: local
operators are reconstructed in terms of the generators of the
so-called Yang-Baxter algebra, which contains in particular  these
creation/annihilation operators for the eigenstates. Step $(ii)$
can then be completed using only the quadratic commutation
relations satisfied by these generators \cite{KitMT00}. The
computation of the resulting scalar products in step $(iii)$ may
also present some technical difficulties. In the case of the $XXZ$
Heisenberg chain, it has been solved using the algebraic structure
of the Yang-Baxter algebra \cite{Sla89,KitMT99}. Finally, step
$(iv)$ is based on the results of \cite{YanY66,YanY66a}.

Note that this procedure remains essentially the same in the case
of the two-point correlation functions (see Section \ref{sec-3}).
The main difference is that, in step $(ii)$, the reconstruction of
the corresponding local operators from the solution of the inverse
problem gives rise to a more complicated combination of the
generators of the Yang-Baxter algebra, so that the use of their
commutation relations to determine their action on the eigenstates
involves a more complicated combinatoric.

\subsection{General framework}\label{sec-21}

To compute the elementary blocks \eq{genabcd00}, or more generally
any correlation function, the first step is to determine the
eigenstates of the Hamiltonian \eq{IHamXXZ} and in particular its
ground state. In the framework of algebraic Bethe Ansatz
\cite{FadST79}, these eigenstates can be described in terms of
generalised creation and annihilation operators which are elements
of the so-called quantum monodromy matrix. In the case of the
$XXZ$ chain \eq{IHamXXZ} the monodromy matrix is  a $2\times 2$
matrix,
\be{ABAT}
  T(\lambda)=\begin{pmatrix}
    A(\lambda) & B(\lambda)\\
    C(\lambda) & D(\lambda)
  \end{pmatrix},
\ee
with operator-valued entries $A,B,C$ and $D$ which depend on a
complex parameter $\lambda$ (spectral parameter) and act in the
quantum space of states ${\cal H}$ of the chain. It is defined as
the ordered product
\be{prod-L}
  T(\lambda)=
      L_{M}(\lambda)\ldots L_{2}(\lambda) L_{1}(\lambda),
\ee
where $L_n(\la)$ denotes the quantum $L$-operator at the site $n$
of the chain:
\be{op-L}
  L_{n}(\lambda)=\begin{pmatrix}
    \sinh(\lambda+\frac{\eta}{2}\, \sigma_n^z)&\sinh\eta \,\sigma_n^-\\
    \sinh\eta\,\sigma_n^+         &\sinh(\lambda-\frac{\eta}{2}\, \sigma_n^z)
  \end{pmatrix}.
\ee
Here and in the following, the parameter $\eta$ is related to the
anisotropy parameter as $\Delta=\ch\eta$.

The quantum operators $A,B,C$ and $D$ satisfy a set of quadratic
commutation relations given by the $R$-matrix of the model, and
generate the so-called Yang-Baxter algebra. These commutation
relations imply in particular that the transfer matrices, defined
as
\be{ABAtr} {\cal T}(\lambda)=\tr T(\lambda)=A(\lambda)+D(\lambda),
\ee
commute for different values of the spectral parameter: $[{\cal
T}(\lambda),{\cal T}(\mu)]=0$. The Hamiltonian \eq{H0} at $h=0$ is
related to ${\cal T}(\lambda)$ by the `trace identity'
\be{ABATI}
  H^{(0)}=2\sinh\eta\
    \frac{d{\cal T}(\lambda)}{d\lambda}
    {\cal T}^{-1}(\lambda)
    \pour{\lambda=\frac{\eta}{2}}-2M\ch\eta.
\ee
Therefore, to diagonalise the Hamiltonian \eq{IHamXXZ}, it is
enough to determine the common eigenstates and eigenvalues of
these transfer matrices.

For technical reasons, it is actually convenient to introduce a
slightly more general object, the twisted transfer matrix
\be{Twis-T-M} {\cal T}_\kappa(\la)=A(\la)+\kappa D(\la), \ee
where $\kappa$ is a complex parameter. The particular case of
${\cal T}_\kappa(\la)$ at $\kappa=1$ corresponds to the usual
(untwisted) transfer matrix ${\cal T}(\la)$. It will be also
convenient to consider an inhomogeneous version of the $XXZ$
chain, for which
\be{prod-Linh}
  T(\lambda)=
      L_{M}(\lambda-\xi_M+\eta/2)\ldots L_{2}(\lambda-\xi_2+\eta/2) \,
      L_{1}(\lambda-\xi_1+\eta/2).
\ee
Here, $\xi_1,\ldots,\xi_M$ are complex parameters (inhomogeneity
parameters) attached to each site of the lattice. The homogeneous
model \eq{IHamXXZ} corresponds to the case where $\xi_j=\eta/2$
for $j=1,\ldots,M$.

In the framework of algebraic Bethe Ansatz, an arbitrary  quantum
state can be obtained from the states generated  by multiple
action of operators $B(\lambda)$ on the reference state $\ket{0}$
with all spins up (respectively by multiple action of operators
$C(\lambda)$ on the dual reference state $\bra{0}$),
\be{ABAES}
  \ket{\psi}=\prod_{j=1}^{N}B(\lambda_j)\ket{0},
  \qquad
  \bra{\psi}=\bra{0}\prod_{j=1}^{N}C(\lambda_j),
  \qquad N=0,1,\dots, M.
\ee
%

\subsection{Description of the eigenstates}\label{sec-22}

Let us consider here the subspace $\mathcal{H}^{(M/2-N)}$ of the
space of states $\mathcal{H}$ with a fixed number $N$ of spins
down. In this subspace, the eigenstates
$\ket{\psi_\kappa(\{\lambda\})}$ (respectively
$\bra{\psi_\kappa(\{\lambda\})}$) of the twisted transfer matrix
${\cal T}_\kappa(\mu)$ can be constructed in the form \eq{ABAES},
where the parameters $\lambda_1,\ldots,\la_N$ satisfy the system
of twisted Bethe equations
\be{TTMBE_Y}
   {\cal Y}_\kappa(\lambda_j|\{\lambda\})=0, \qquad j=1,\dots,N.
\ee
Here, the function ${\cal Y}_\kappa$ is defined as
\be{TTM_Y-def}
   {\cal Y}_\kappa(\mu|\{\lambda\}) =
      a(\mu)\prod_{k=1}^{N}\sinh(\lambda_k-\mu+\eta)
      + \kappa\,d(\mu) \prod_{k=1}^{N}\sinh(\lambda_k-\mu-\eta),
\ee
and $a(\lambda)$, $d(\lambda)$ are the eigenvalues of the
operators $A(\lambda)$ and $D(\lambda)$ on the reference state
$\ket{0}$. In the normalisation \eq{op-L} and for the
inhomogeneous model \eq{prod-Linh}, we have
\begin{equation}\label{EV_DA}
  a(\lambda) =\prod_{a=1}^M\sinh(\lambda-\xi_a+\eta),\qquad
  d(\lambda) =\prod_{a=1}^M\sinh(\lambda-\xi_a).
\end{equation}
The corresponding eigenvalue of ${\cal T}_\kappa(\mu)$ on
$\ket{\psi_\kappa(\{\lambda\})}$ (or on a dual eigenstate) is
\be{ABAEV} \tau_\kappa(\mu|\{\lambda\})=
a(\mu)\prod_{k=1}^{N}\frac{\sinh(\lambda_k-\mu+\eta)}{\sinh(\lambda_k-\mu)}
+ \kappa\,d(\mu)
\prod_{k=1}^{N}\frac{\sinh(\mu-\lambda_k+\eta)}{\sinh(\mu-\lambda_k)}.
\ee

The solutions of the system of twisted Bethe equations
\eq{TTMBE_Y} have been analysed in \cite{TarV96}. In general, not
all of these solutions correspond to eigenvectors of ${\cal
T}_\kappa(\mu)$.
\begin{Def}
   A solution $\{\lambda\}$ of the system \eq{TTMBE_Y} is called
   {\sl admissible} if
\be{admiss}
  d(\lambda_j)
  \prod_{\substack{k=1\\k\ne j}}^N\sinh(\lambda_j-\lambda_k+\eta)\ne0,\qquad
  j=1,\dots,N,
\ee
  and {\sl unadmissible} otherwise.
  A solution is called {\sl off-diagonal} if the corresponding parameters
  $\lambda_1,\dots,\lambda_N$ are pairwise distinct,
  and {\sl diagonal} otherwise.
\end{Def}
One of the main result of \cite{TarV96} is that, for generic
parameters $\kappa$ and $\{\xi\}$, the set of the eigenstates
corresponding to the admissible off-diagonal solutions of the
system of twisted Bethe equations \eq{TTMBE_Y}  form a basis in
the subspace ${\cal H}^{(M/2-N)}$. It has been proven in
\cite{KitMST04b} that this result is still valid in the
homogeneous case $\xi_j=\eta/2$, $j=1,\ldots,N$, at least if
$\kappa$ is in a punctured vicinity of the origin (i.e.
$0<|\kappa|<\kappa_0$ for $\kappa_0$ small enough). Note however
that, for specific values of $\kappa$ and $\{\xi\}$, the basis of
the eigenstates in ${\cal H}^{(M/2-N)}$ may include some states
corresponding to unadmissible solutions of \eq{TTMBE_Y} (in
particular in the homogeneous limit at $\kappa=1$).

At $\kappa=1$, it follows from the trace identity \eq{ABATI} that
the eigenstates of the transfer matrix  coincide, in the
homogeneous limit, with the ones of the Hamiltonian \eq{IHamXXZ}.
The corresponding eigenvalues in the case of zero magnetic field
can be obtained from \eq{ABATI}, \eq{ABAEV}:
\be{eigen-val-H}
  H^{(0)}\, \ket{\psi(\{\lambda\})}=\sum_{j=1}^N E(\lambda_j)\cdot
    \ket{\psi(\{\lambda\})},
\ee
where the bare one-particle energy $E(\lambda)$ is equal to
\be{E}
  E(\lambda)=\frac{2\sinh^2\eta}
                  {\sinh(\lambda+\frac{\eta}{2})\sinh(\lambda-\frac{\eta}{2})}.
\ee
One can similarly define the bare one-particle momentum. It is
given by
\be{P}
  p(\lambda)=i\log\left(\frac{\sinh(\lambda-\frac{\eta}{2})}
                             {\sinh(\lambda+\frac{\eta}{2})}\right).
\ee
%

\subsection{Action of local operators on eigenstates}\label{sec-23}

A local operator $E^{\epsilon'_j,\epsilon_j}_{j}$, acting in a
local quantum space ${\cal H}_j$ at site $j$, can also be
expressed in terms of the entries of the monodromy matrix by
solving the quantum inverse scattering problem
\cite{KitMT99,MaiT00}:
\be{FCtab}
  E^{\epsilon'_j,\epsilon_j}_{j}
  =\prod_{\alpha=1}^{j-1}{\cal T}(\xi_\alpha)\cdot
    T_{\epsilon_j,\epsilon'_j}(\xi_j)
   \cdot\prod_{\alpha=1}^{j}{\cal T}^{-1}(\xi_\alpha).
\ee

This enables us to use the quadratic commutation relations for the
generators $A, B, C, D$ of the Yang-Baxter algebra to get the
action of any product of local operators on arbitrary states of
the form \eq{ABAES} \cite{KitMT00}:
\be{action}
   \bra{0}\pl_{k=1}^{N}C(\lambda_k)\cdot
      \pl_{j=1}^{m}T_{\epsilon_j,\epsilon'_j}(\lambda_{N+j})
   =
  \sul_{\mathcal{P}\subset\{\lambda\}}
          \mathcal{F}_{\mathcal{P}}(\{\lambda\},\{\epsilon_j,\epsilon'_j\})\
   \bra{0} \pl_{b\in\mathcal{P}} C(\la_b),
\ee
in which the sum is taken over subsets $\mathcal{P}$ of cardinal
$N$ of the set $\{\la_1,\ldots,\la_{N+m}\}$, and the coefficients
$\mathcal{F}_{\mathcal{P}}(\{\lambda\},\{\epsilon_j,\epsilon'_j\})$
can be computed generically. Thus, the elementary blocks
\eq{genabcd00}, and more generally any correlation functions (see
Section~\ref{sec-3}), can be expressed as some sums over scalar
products of a Bethe state with an arbitrary state of the form
\eq{ABAES}.

\subsection{Scalar products}\label{sec-24}

We recall here the expressions for the scalar product of an
eigenstate  of the twisted transfer matrix with any arbitrary
state of the form \eq{ABAES}.

Let us first define, for arbitrary positive integers $n,n'$ ($n\le
n'$) and
arbitrary sets of variables $\lambda_1,\dots,\lambda_n$, $\mu_1,%
\dots,\mu_n$ and $\nu_1,\dots,\nu_{n'}$ such that $\{\lambda\}
\subset \{\nu\}$,  the $n\times n$ matrix
$\Omega_\kappa(\{\lambda\},\{\mu\}|\{\nu\})$  as
\begin{multline} \label{matH}
  (\Omega_\kappa)_{jk}(\{\lambda\},\{\mu\}|\{\nu\})=
  a(\mu_k)\,t(\lambda_j,\mu_k)\,\prod_{a=1}^{n'} \sinh(\nu_a-\mu_k+\eta)\\
   -\kappa\, d(\mu_k)\,t(\mu_k,\lambda_j)\,\prod_{a=1}^{n'} \sinh(\nu_a-\mu_k-\eta),
\end{multline}
with
\be{def-t}
t(\lambda,\mu)=\frac{\sinh\eta}{\sinh(\lambda-\mu)\sinh(\lambda-\mu+\eta)}.
\ee

\begin{prop}\textup{ \cite{Sla89,KitMT99,KitMST04a}}
Let $\{\lambda_1,\dots,\lambda_N\}$ be a solution of the system of
twisted Bethe equations \eq{TTMBE_Y}, and $\mu_1,\dots,\mu_N$ be
generic complex numbers. Then
\begin{align}
  \bra{0}\prod_{j=1}^{N}C(\mu_j)\,\ket{\psi_\kappa(\{\lambda\})}
  &=\bra{\psi_\kappa(\{\lambda\})}\prod_{j=1}^{N}B(\mu_j)\ket{0}
        \nonumber\\
  &=\frac{ \pl_{a=1}^{N} d(\lambda_a)
           \prod\limits_{a,b=1}^N\sinh(\mu_b-\lambda_a)}
         {\prod\limits_{a>b}^N\sinh(\lambda_a-\lambda_b)\sinh(\mu_b-\mu_a)}
      \cdot
   \det_N \left(\frac\partial{\partial\lambda_j}
   \tau_\kappa(\mu_k|\{\lambda\})\right)\label{FCdet}
               \\
  &= \frac{\pl_{a=1}^{N} d(\lambda_a)}
          {\prod\limits_{a>b}^N\sinh(\lambda_a-\lambda_b)\sinh(\mu_b-\mu_a)}
     \cdot
     \det_N \Omega_\kappa(\{\lambda\},\{\mu\} | \{\lambda\})\label{scal-prod}.
\end{align}
\end{prop}

\begin{rem}
If the sets $\{\lambda\}$ and $\{\mu\}$ are different, the
eigenstate $\ket{\psi_\kappa(\{\lambda\})}$ is orthogonal to the
dual eigenstate $\bra{\psi_\kappa(\{\mu\})}$. Otherwise
\begin{align}\label{Y-Jac}
  \bracket{\psi_\kappa(\{\lambda\})\,|\,\psi_\kappa(\{\lambda\})}
  &=\frac{\pl_{a=1}^{N} d(\lambda_a)}
         {\prod\limits_{\substack{a,b=1 \\ a\ne b}}^N\sinh(\lambda_a-\lambda_b)}
    \cdot \det_N \Omega_\kappa(\{\lambda\},\{\lambda\}|\{\lambda\})
        \\
  &=(-1)^N\frac{\pl_{a=1}^{N} d(\lambda_a)}
               {\pl_{\substack{a,b=1 \\ a\ne b}}^N\sinh(\lambda_a-\lambda_b)}
    \cdot \det_N \left(\frac\partial{\partial\lambda_k}
                {\cal Y}_\kappa(\lambda_j|\{\lambda\})\right).
\label{norm}
\end{align}
\end{rem}

The equations \eq{FCdet}--\eq{norm} are valid for any arbitrary
complex parameter $\kappa$, in particular at $\kappa=1$. In this
case we may omit the subscript $\kappa$ and denote
$(\psi,\tau,\mathcal{Y},\Omega)=\left.
 (\psi_\kappa,\tau_\kappa,\mathcal{Y}_\kappa,\Omega_\kappa)\right|_{\kappa=1}$.

Using these expressions for the scalar product and the norm of the
Bethe state, one sees from equation \eq{action} that the
correlation functions can be expressed as (multiple) sums of
determinants \cite{KitMT00}.

\subsection{Elementary blocks in the thermodynamic limit}\label{sec-25}

In the thermodynamic limit, the system of Bethe equations for the
ground state turns into a single integral equation for the ground
state spectral density $\rho_{\text{tot}}(\lambda)$
\cite{YanY66a}:
\begin{equation}\label{GFLiebeq}
  \rho_{\text{tot}}(\lambda)
  +\int\limits_{\cal C} K(\lambda-\mu)\,\rho_\text{tot}(\mu)\,d\mu
  = \frac{i}{2\pi}t(\lambda,\eta/2),
\end{equation}
where the contour ${\cal C}$, which depends on the value of the
magnetic field $h$, is an interval of the real axis in the
massless regime and of the imaginary axis in the massive regime
(${\cal C} = [-\Lambda_h,\Lambda_h]$), and the kernel $K$ is given
by
\be{GFKkern}
K(\lambda)=\frac{i\sinh2\eta}{2\pi\sinh(\lambda+\eta)\sinh(\lambda-\eta)}.
\ee
For technical convenience, one can also define an inhomogeneous
version $\rho(\lambda,\xi)$ of this ground state density as the
solution of the equation
\begin{equation}\label{GFinteqinh}
  \rho(\lambda,\xi)
  +\int\limits_{\cal C} K(\lambda-\mu)\,\rho(\mu,\xi)\,d\mu
  = \frac{i}{2\pi}t(\lambda,\xi).
\end{equation}
Note that $\rho_\text{tot}(\lambda)=\rho(\lambda,\eta/2)$. In the
case of zero magnetic field, this integral equation can be solved
explicitely and we have
\begin{xalignat}{3}
 &|\Delta|<1: & &\Lambda_h =\Lambda= \infty, \label{inf1}& & \\
 & &  &\rho(\lambda,\xi) =\frac{i}{2\zeta\sinh\frac\pi\zeta(\lambda-\xi)},
         \quad
    & &(\zeta=i\eta>0),\\
 &\Delta>1: & &\Lambda_h =\Lambda= -i\pi/2, \label{inf2}& & \\
 & &  &\rho(\lambda,\xi) =-\frac{1}{2\pi}
      \prod_{n=1}^\infty\left(\frac{1-q^{2n}}{1+q^{2n}}\right)^2
      \frac{\vartheta_2(i(\lambda-\xi),q)}{\vartheta_1(i(\lambda-\xi),q)},
    & &(\zeta=-\eta>0, \ q=e^\eta).
\end{xalignat}

More generally, in the limit $M\to\infty$, sums over the solutions
$\la_1,\ldots,\la_N$ of Bethe equations for the ground state
become integrals over the density $\rho_{\text{tot}}$:
\be{sum-int}
  \frac 1M\sul_{j=1}^{N} f(\la_j)
     =\int\limits_{\cal C}\rho_\text{tot}(\la)f(\la)\,d\la+o(1/M),
\ee
for any smooth bounded function $f(\la)$. This leads to a multiple
integral representation for the correlation functions. In
particular, the $m$-point elementary blocks \eq{genabcd00} can be
written as a $m$-fold multiple integral of the form \cite{KitMT00}
\be{resultEB}
   F_m(\{\epsilon_j,\epsilon'_j\})=
     \bigg(\prod_{j=1}^m\int\limits_{{\cal C}_j}d\la_j\bigg)\,
     \mathcal{F}(\{\la\},\{\epsilon_j,\epsilon'_j\})\,
     \mathcal{S}(\{\la\}).
\ee
In this expression, the set of integration contours $\{{\cal C}_j,
j=1,\ldots,m\}$ depends on the regime, on the value of the
magnetic field, and on the configuration
$\{\epsilon_j,\epsilon'_j\}$ of the block we consider (see
\cite{KitMT00}). The integrand can be split into two parts: a
purely algebraic quantity
$\mathcal{F}(\{\la\},\{\epsilon_j,\epsilon'_j\})$ which arises
from the commutation relation of the monodromy matrix elements and
does not depend on the ground state, and a quantity
$\mathcal{S}(\{\la\})$ which is the same for all blocks and
contains all the informations about the ground state. The latter
is actually a functional of the ground state density that comes
from the thermodynamic limit of the normalised scalar product. In
the general inhomogeneous case, it is given as
\be{densityM}
  \mathcal{S}(\{\la\})=\pl_{1\le j<k\le m}\frac 1{\sinh(\xi_j-\xi_k)}\cdot
                            \det_{1\le j,k\le m}[\rho(\la_j,\xi_k)].
\ee
We refer to \cite{KitMT00} for an explicit expression of the
algebraic part $\mathcal{F}(\{\la\},\{\epsilon_j,\epsilon'_j\})$.
Let us just mention here that one can essentially distinguish two
types of integrals at this level: what we will call `$D$-type'
integral, that comes from the contribution of the action of an
operator $D$ on a state of the from \eq{ABAES}, with its
corresponding algebraic part and integration contour
$\mathcal{C}_j=\mathcal{C}$, and the `$A$-type' integral,
associated to the action of operator $A$, with a different
algebraic part and a contour $\mathcal{C}_j$ which is shifted
compared to the contour $\mathcal{C}$ of the integral equation
\eq{GFLiebeq} for the ground state density.  As the action of
operator $B$ is very similar to the successive action of operators
$A$ and $D$, it produces in the final result both types of
integrals.

Let us finally note that the representation \eq{resultEB} for the
elementary block \eq{genabcd00} coincides exactly for zero
magnetic field with the multiple integral representation obtained
and conjectured in \cite{JimMMN92,JimM96} (see also
\cite{JimM95L}) from the $q$-vertex operator approach, and
generalises it to the case of a non-zero magnetic field for which
the quantum affine symmetry used in \cite{JimM95L} is broken.


\section{Re-summation of the elementary blocks\label{sec-3}}

The method presented in the last section is quite straightforward
and gives formally the possibility to compute any  correlation
function. However, it has been developed for the computation of
the expectation values of the monomials $T_{a_1b_1}(\xi_1)\cdots
T_{a_mb_m}(\xi_m)$, leading to the evaluation of elementary
building blocks, whereas the study of the two-point functions
involves big sums of such monomials.

Indeed, let us consider for example the correlation function
$\bracket{\sigma_1^+\,\sigma_{m+1}^-}$. Then, according to the
solution of the inverse scattering problem \eq{FCtab}, we need to
calculate the expectation value
\be{CCF+-}
  \bra{\psi(\{\lambda\})}\,C(\xi_1)\cdot
          \prod_{a=2}^{m}{\cal T}(\xi_a)\cdot
          B(\xi_{m+1})\cdot \prod_{b=1}^{m+1}{\cal T}^{-1}(\xi_b)\,
\ket{\psi(\{\lambda\})}. \ee
Since $\ket{\psi(\{\lambda\})}$ is an eigenstate of the transfer
matrix ${\cal T}$, the action of $\prod_{b=1}^{m+1} {\cal
T}^{-1}(\xi_b)$ on this state merely produces a numerical factor.
However, it is much more complicated to evaluate the action of
$\prod_{a=2}^{m}{\cal T}(\xi_a)$. Indeed, we have to act first
with $C(\xi_1)$ on $\bra{\psi(\{\lambda\})}$ (or with
$B(\xi_{m+1})$ on $\ket{\psi(\{\lambda\})}$), which gives  a sum
of states which are no longer eigenstates of the transfer matrix,
and on which the multiple action of ${\cal T}$ is not simple. In
fact,  in the framework of the approach of Section~\ref{sec-2},
the product $\prod_{a=2}^{m}(A+D)(\xi_a)$ would be computed as a
sum of $2^{m-1}$ monomials, which eventually would lead to a huge
sum of elementary blocks. This is not very convenient, in
particular at large distance $m$. Therefore, to obtain manageable
expressions for such correlation functions, it is of great
importance to develop an alternative and compact way to express
the multiple action of the transfer matrix on arbitrary states or,
in other words, to make an effective re-summation of the
corresponding sum of $2^{m-1}$ terms.

In this section, we explain two different ways to perform such
re-summations.

\subsection{Re-summation with auxiliary integrals}\label{sec-31}

The results presented in this subsection were first  obtained in
\cite{KitMST02a}. We recall here the main steps of this
re-summation.

Let us consider the multiple action of the twisted transfer
matrices on an arbitrary dual state
$\bra{0}\prod_{j=1}^NC(\mu_j)$,
\be{mult-act}
  \bra{0}\prod_{j=1}^NC(\mu_j)\prod_{a=1}^m{\cal T}_\kappa(x_a),
\ee
where $x_1,\dots,x_m$ and $\mu_1,\dots,\mu_N$ are generic complex
numbers. Using the quadratic commutation relations between $A$,
$D$ and $C$, one can prove

\begin{prop}\label{prop1-actTM}
{\rm{ \cite{KitMST02a} }} Let $\kappa$, $x_1,\ldots,x_m$ and
$\mu_1,\dots,\mu_N$ be generic complex numbers. The action of
$\prod_{a=1}^m{\cal T}_\kappa(x_a)$ on the state
$\bra{0}\prod_{j=1}^NC(\mu_j)$ can be written as
 \begin{multline}\label{APcompl-act}
    \bra{0}\prod_{j=1}^N C(\mu_j) \prod_{a=1}^m{\cal T}_\kappa(x_a)
      \\
 \Negspace
   =\sum_{n=0}^{\min{(m,N)}} \!\!
    \sum_{\substack{
           \{\mu\}=\{\mu_{\alpha_+}\}\cup\{\mu_{\alpha_-}\}\\
           \{x\}=\{x_{\gamma_+}\}\cup\{x_{\gamma_-}\}\\
           |\alpha_+|=|\gamma_+|=n}} \Negspace
     R_n^\kappa(\{x_{\gamma_+}\},\{x_{\gamma_-}\},
                \{\mu_{\alpha_+}\},\{\mu_{\alpha_-}\})\,
     \bra{0}\prod_{a\in\gamma_+} C(x_a) \prod_{b\in\alpha_-}C(\mu_b).
\end{multline}
In this expression the set of parameters $\{\mu\}$ is divided into
two subsets $\{\mu\}=\{\mu_{\alpha_+}\}\cup\{\mu_{\alpha_-}\}$
such that $\{\mu_{\alpha_+}\}\cap\{\mu_{\alpha_-}\}=\emptyset$.
Similarly the set $\{x\}$ is also  divided as
$\{x\}=\{x_{\gamma_+}\}\cup\{x_{\gamma_-}\}$,
$\{x_{\gamma_+}\}\cap\{x_{\gamma_-}\}=\emptyset$. These partitions
are independent except that $\#\{x_{\gamma_+}\}=
\#\{\mu_{\alpha_+}\}=n$. The sum in \eq{APcompl-act} is taken with
respect to all such partitions, and the corresponding coefficient
$R_n^\kappa(\{x_{\gamma_+}\},\{x_{\gamma_-}\},
                \{\mu_{\alpha_+}\},\{\mu_{\alpha_-}\})$ is given by
 \begin{multline}\label{APRn}
 R_n^\kappa =
    \biggl\{
      \prod_{\substack{a>b \\ a,b\in\alpha_+}}\sinh(\mu_b-\mu_a)
      \prod_{\substack{a<b \\ a,b\in\gamma_+}}\sinh(x_b-x_a)
      \prod_{a\in\alpha_+}\prod_{b\in\alpha_-}\sinh(\mu_b-\mu_a)
    \biggr\}^{-1}\\
 \times
    \prod_{a\in\gamma_-}\tau_\kappa(x_a|\{x_{\gamma_+}\}\cup
                \{\mu_{\alpha_-}\})  \cdot
    \det_n \Omega_\kappa(\{x_{\gamma_+}\},\{\mu_{\alpha_+}\}\mid
                        \{x_{\gamma_+}\}\cup\{\mu_{\alpha_-}\}).
 \end{multline}
\end{prop}

The equations \eq{APcompl-act}, \eq{APRn} are the key-formulae of
our re-summation. When applying these expressions to particular
cases, one obtains directly new multiple integral representations
for the two-point correlation functions which are essentially
different from the ones that result from  the elementary blocks
approach.

\bigskip

One of the simplest applications of Proposition~\ref{prop1-actTM}
concerns the generating function of the two-point correlation
function of the third components of spin, which is defined as the
expectation value
\be{def-Q}
 \bracket{Q^\kappa_{l,m}}= \frac{\bra{\psi(\{\lambda\})}\,Q^\kappa_{l,m}\,
                                                    \ket{\psi(\{\lambda\})}}
                          {\bracket{\psi(\{\lambda\})\mid\psi(\{\lambda\})}}
 \ee
of the operator
\be{GFdefQ}
  Q^\kappa_{l,m}
  =\prod_{n=l}^m\left(\frac{1+\kappa}2
                        +\frac{1-\kappa}2\cdot\sigma_n^z\right)
  =\prod_{j=1}^{l-1}{\cal T}(\xi_j)\cdot \prod_{j=l}^m{\cal T}_\kappa(\xi_j)
              \cdot\prod_{j=1}^m{\cal T}^{-1}(\xi_j),
\ee
where $\ket{\psi(\{\lambda\})}$ is an eigenstate of
$\mathcal{T}(\mu)$ in the subspace $\mathcal{H}^{(M/2-N)}$. The
two-point correlation function of the third components of local
spins in the eigenstate $\ket{\psi(\{\lambda\})}$   can be
obtained in terms of the second `lattice derivative' and the
second derivative with respect to $\kappa$ of the generating
function \eq{def-Q} at $\kappa=1$:
\begin{multline}\label{cor-funct-ss}
  \bracket{\sigma_{l}^z\,\sigma_{l+m}^z}
   =\bracket{\sigma_{l}^z}+\bracket{\sigma_{l+m}^z}-1\\
    +
     2\frac{\partial^2}{\partial\kappa^2}
        \bracket{ Q^\kappa_{l,l+m} - Q^\kappa_{l,l+m-1} -
          Q^\kappa_{l+1,l+m} + Q^\kappa_{l+1,l+m-1}}\pour{\kappa=1}.
\end{multline}
Due to the translational invariance of the correlation functions
in the homogeneous model, we will simply consider the following
expectation value:
\be{exp-Q}
 \bracket{Q^\kappa_{1,m}}
 = \prod_{j=1}^m \tau^{-1}(\xi_j|\{\lambda\})\cdot
 \frac{\bra{\psi(\{\lambda\})}\prod\limits_{j=1}^m{\cal T}_\kappa(\xi_j)\,
                                                    \ket{\psi(\{\lambda\})}}
      {\bracket{\psi(\{\lambda\})\mid\psi(\{\lambda\})}}.
 \ee

In order to evaluate this generating function, one should first
compute the multiple action of ${\cal T}_\kappa(\xi_j)$ in the
r.h.s. of \eq{exp-Q} by means of Proposition~\ref{prop1-actTM},
and then project the result on the eigenstate
$\ket{\psi(\{\lambda\})}$ using \eq{FCdet} for the scalar product.
Hereby the expression of the coefficient $R_n^\kappa$ and of the
matrices $\Omega$, $\Omega_\kappa$ can be simplified  using Bethe
equations for the set $\{\lambda\}$ and the fact that
$d(\xi_j)=0$. Note also that we can restrict ourselves to the case
$m<N$, since eventually we are going to compute the correlation
function in the thermodynamic limit. The result can be written in
the following form
\be{Q-interm}
 \bracket{Q^\kappa_{1,m}}=
 \sum_{n=0}^{m}
    \sum_{\substack{
           \{\lambda\}=\{\lambda_{\alpha_+}\}\cup\{\lambda_{\alpha_-}\}\\
           \{\xi\}=\{\xi_{\gamma_+}\}\cup\{\xi_{\gamma_-}\}\\
           |\alpha_+|=|\gamma_+|=n}}\prod_{a\in\gamma_-}\prod_{b\in\gamma_+}
\frac{\sinh(\xi_b-\xi_a+\eta)}{\sinh(\xi_b-\xi_a)}
     \cdot{\cal F}_n^\kappa(\{x_{\gamma_+}\},
                \{\lambda_{\alpha_+}\},\{\lambda_{\alpha_-}\}).
\ee
Here we have combined all the factors in one function ${\cal F}_n$
and extracted explicitly the dependency on the subset
$\{\xi_{\gamma_-}\}$. We refer to \cite{KitMST02a} for a more
explicit expression.

Let us now suppose that $\ket{\psi(\{\la\})}$ is the eigenstate of
the inhomogeneous transfer matrix which tends, in the homogeneous
limit, toward the ground state of the Hamiltonian \eq{IHamXXZ}.
Then, in the thermodynamic limit,  the sum over the partitions of
the set $\{\lambda\}$ turns, for each given $n$, into an $n$-fold
multiple integral over the support of the ground state density,
just  like in Section~\ref{sec-25} for the elementary blocks. As
for the sum over the partitions of the set $\{\xi\}$, it can be
computed in terms of some auxiliary contour integrals. Indeed, it
is easy to see that
\begin{multline}\label{GFcontint}
 \sum_{\substack{
 \{\xi\}=\{\xi_{\gamma_-}\}\cup\{\xi_{\gamma_+}\} \\ |\gamma_+|=n }}
 \prod_{a\in\gamma_-}\prod_{b\in\gamma_+}
 \frac{\sinh(\xi_b-\xi_a+\eta)}{\sinh(\xi_b-\xi_a)}
     \cdot{\cal F}_n^\kappa(\{x_{\gamma_+}\},
                \{\lambda_{\alpha_+}\},\{\lambda_{\alpha_-}\})\\
 %
 =\frac1{n!}\oint\limits_{\Gamma\{\xi\} } \prod_{j=1}^n\frac{dz}{2\pi i} \,
    \prod_{a=1}^n\prod_{b=1}^{m}
     \frac{\sinh(z_a-\xi_b+\eta)}{\sinh(z_a-\xi_b)}\\
 \times
 \frac{\prod\limits_{a=1}^n\prod\limits_{\substack{b=1\\b\ne a}}^n
                                    \sinh(z_a-z_b)}
      {\prod\limits_{a=1}^n\prod\limits_{b=1}^n \sinh(z_a-z_b+\eta)}\cdot
 {\cal
 F}_n^\kappa(\{z\},\{\lambda_{\alpha_+}\},\{\lambda_{\alpha_-}\}).
\end{multline}
Here the contour $\Gamma\{\xi\}$ surrounds the points
$\xi_1,\dots, \xi_{m}$ and does not contain any other
singularities of the integrand. Observe that this representation
allows one to take the homogeneous limit directly by setting
$\xi_j=\eta/2$ in the expression.

Thus, the sum over partitions in \eq{Q-interm} can be written in
terms of multiple integrals. The resulting representation for the
generating function of the correlation function
$\bracket{\sigma_{1}^z\,\sigma_{m+1}^z}$ has the following form
\cite{KitMST02a}:
\begin{multline}\label{Q-int-rep}
 \bracket{Q^\kappa_{1,m}}=
 \sum_{n=0}^{m} \frac1{(n!)^2}
         \oint\limits_{\Gamma\{\xi\}} \prod_{j=1}^n\frac{dz}{2\pi i}
         \int\limits_{\cal C}d^n\lambda\,
 \prod_{a=1}^n\prod_{b=1}^{m}
 \frac{\sinh(z_a-\xi_b+\eta)\sinh(\lambda_a-\xi_b)}
      {\sinh(z_a-\xi_b)\sinh(\lambda_a-\xi_b+\eta)}
                 \\
 \times W_n(\{\lambda\},\{z\})\cdot\det_n M_\kappa(\{\lambda\},\{z\})\cdot
 \det_n\rho(\lambda_j,z_k),
 \end{multline}
with
\be{W}
 W_n(\{\lambda\},\{z\})=\prod_{a=1}^n\prod_{b=1}^{n}
 \frac{\sinh(z_a-\lambda_b+\eta)\sinh(\lambda_b-z_a+\eta)}
 {\sinh(z_a-z_b+\eta)\sinh(\lambda_a-\lambda_b+\eta)},
\ee
and
\be{M-kappa}
 (M_\kappa)_{jk}(\{\lambda\},\{z\})=t(z_k,\lambda_j)+\kappa \,t(\lambda_j,z_k)
\prod_{a=1}^n
 \frac{\sinh(\lambda_a-\lambda_j+\eta)\sinh(\lambda_j-z_a+\eta)}
 {\sinh(\lambda_j-\lambda_a+\eta)\sinh(z_a-\lambda_j+\eta)}.
\ee
The integration contour ${\cal C}$ and the density function
$\rho(\lambda,z)$ are defined in \eq{GFLiebeq}--\eq{GFinteqinh}.

If we had used the expressions of the elementary blocks derived in
Section~\ref{sec-2}, we would have obtained the generating
function $\bracket{Q^\kappa_{1,m}}$ as a sum of $2^m$ terms, each
of them being written as a $m$-multiple integral of the type
 \eq{resultEB}. Instead,
we have now a representation containing only $m$ nontrivial terms.
The $n$-th term is formulated as a $2n$-fold multiple integral,
with $n$ integrals
 over the support of the ground state density  and $n$
auxiliary contour integrals over some auxiliary variables $z_j$.
We will see in Section~\ref{sec-5} that these last integrals play
the role of
 an effective re-summation of the form factor series.

Observe also that, in the homogeneous model, the dependency on the
distance $m$ enters each integral only as a power of a simple
function. This fact might be used for the asymptotic analysis of
these multiple integrals by the steepest descent  method.

\bigskip

Other two-point functions can be considered in a similar manner.
For example, the expectation value \eq{CCF+-} gives us the
correlation function $\bracket{\sigma_1^+\sigma_{m+1}^-}$. It is
clear that one can evaluate this correlation function by using
first the equations \eq{APcompl-act}, \eq{APRn} of
Proposition~\ref{prop1-actTM}, by acting in a second step with the
operator $B(\xi_{m+1})$ on the resulting states, and by finally
computing the corresponding scalar products via \eq{FCdet}. All
the steps of this derivation are quite similar to the ones that we
have just described in the case of the generating function
$\bracket{Q^\kappa_{1,m}}$. Let us merely give here the new
multiple integral representation that we obtain by this method for
the ground-state correlation function
$\bracket{\sigma_1^+\,\sigma_{m+1}^-}$ in the thermodynamic limit.
For simplicity, we present the answer in the homogeneous limit and
at zero magnetic field:
\begin{multline}\label{Os+s-}
 \bracket{\sigma^+_1\sigma^-_{m+1}}=\sum_{n=0}^{m-1}
  \frac1{n!(n+1)!}
  \oint\limits_{\Gamma\{\frac\eta2\}} \prod_{j=1}^{n+1}\frac{dz_j}{2\pi i}
  \int\limits_{\cal C} d^{n+2}\lambda
  \left(
    \prod_{a=1}^{n+1}\frac{\sinh(z_a+\frac\eta2)}{\sinh(z_a-\frac\eta2)}
   \cdot \prod_{a=1}^{n} \frac{\sinh(\lambda_a-\frac\eta2)}
                              {\sinh(\lambda_a+\frac\eta2)}
         \right)^{\!\!\! m} \\
\times \frac1{\sinh(\lambda_{n+1}-\lambda_{n+2})}\cdot
\left(\frac{\prod\limits_{a=1}^{n+1}
\sinh(\lambda_{n+1}-z_a+\eta)\sinh(\lambda_{n+2}-z_a)}
{\prod\limits_{a=1}^n\sinh(\lambda_{n+1}-\lambda_a+\eta)
\sinh(\lambda_{n+2}-\lambda_a)} \right)\cdot \hat W_n(\{\lambda\},\{z\})\\
 \times \det_{n+1} \hat M_\kappa(\{\lambda\},\{z\})\cdot
\det_{n+2}\left[
\rho(\lambda_j,z_1),\dots,\rho(\lambda_j,z_{n+1}),
\rho(\lambda_j,{\textstyle\frac{\eta}{2}}) \right],
\end{multline}
where the contours ${\cal C}$ and $\Gamma\{\frac\eta2\}$ are the
same as in \eq{Q-int-rep}. The analogue
$\hat{W}_n(\{\lambda\},\{z\})$ of the function
$W_n(\{\lambda\},\{z\})$ is
\be{OhWn} \hat W_n(\{\lambda\},\{z\})=
\frac{\prod\limits_{a=1}^n\prod\limits_{b=1}^{n+1}
\sinh(\lambda_a-z_b+\eta)\sinh(z_b-\lambda_a+\eta)}
{\prod\limits_{a=1}^n\prod\limits_{b=1}^n
\sinh(\lambda_a-\lambda_b+\eta)
\prod\limits_{a=1}^{n+1}\prod\limits_{b=1}^{n+1}
\sinh(z_a-z_b+\eta)}, \ee
and the $(n+1)\times(n+1)$ matrix $\hat M_\kappa$ has the entries
\begin{align}\label{OhM}
  &(\hat M_\kappa)_{jk}=t(z_k,\lambda_j)-t(\lambda_j,z_k)
    \prod_{a=1}^n\frac{\sinh(\lambda_a-\lambda_j+\eta)}
{\sinh(\lambda_j-\lambda_a+\eta)}
\prod_{b=1}^{n+1}\frac{\sinh(\lambda_j-z_b+\eta)}
{\sinh(z_b-\lambda_j+\eta)},
\quad j\le n,\\
  &(\hat M_\kappa)_{n+1,k}=t(z_k,{\textstyle \frac\eta2}).
\end{align}

\subsection{Alternative method}

There exists another way to reduce the number of terms in the
multiple integral representations for  the two-point functions. In
fact, the re-summation which has just been described has been
performed at the algebraic level: we have computed algebraically
the multiple action of the twisted transfer matrices on an
arbitrary state and, thus, we have avoided any mention of the
elementary blocks. On the contrary, the method that will be
presented below deals directly with the elementary blocks in the
thermodynamic limit.

\bigskip

Let us consider again the generating function
$\bracket{Q^\kappa_{1,m}}$ for the correlation function of the
third components of spin. It has been already mentioned  that one
can, in the multiple integral representations \eq{resultEB} for
the elementary blocks, distinguish two types of integrals: the
`$D$-type' integrals (with the original contour $\mathcal{C}$),
and the `$A$-type' integrals (with a shifted contour). In fact,
the generating function \eq{exp-Q} can be decomposed as a sum over
elementary blocks obtained as expectation values of products of
operators $A$ and $D$ only. Such elementary blocks, containing
only diagonal elementary matrices, can be in general written in
the following form:
\begin{multline}
 F_m(\{\epsilon_j,\epsilon_j\})
   =
  \int\limits_{\cal C}d\la_1\dots
  \int\limits_{\cal C}d\la_m\,\,
  \mathcal{S}(\{\la\})
     \\
 \times
\pl_{j>k}{\frac{\sinh(\la_j-\xi_k+(\epsilon_j-1)\eta)\,
          \sinh(\la_k-\xi_j+(2-\epsilon_k)\eta)}
         {\sinh(\la_j-\la_k-(3-\epsilon_j-\epsilon_k)\eta)}}
, \label{EB-2S}
\end{multline}
where the indexes $\epsilon_j$ can take two values $1$ or $2$:
$\epsilon_j=1$ corresponds to an `$A$-type' integral and
$\epsilon_j=2$ corresponds to a `$D$-type' integral. For
simplicity reason, we consider here only the zero magnetic field
case, but a representation similar to \eq{EB-2S}, with more
complicated integration contours, can also be written in the case
of a non-zero external magnetic field. Hence, the generating
function $\bracket{Q^\kappa_{1,m}}$ can be expressed as a sum of
$2^m$ such terms. It is easy to see that the terms which have the
same number of `$A$-type' integrals exhibit a quite similar
structure. This observation permits to write the generating
function $\bracket{Q^\kappa_{1,m}}$ as power series on $\kappa$,
\begin{equation}
  \bracket{Q^\kappa_{1,m}}=\sul_{s=0}^m \kappa^s G_s(m),
\end{equation}
where the coefficient $G_s(m)$ collects all the terms containing
$s$ `$D$-type' integrals and $m-s$ `$A$-type' integrals. This
coefficient can be expressed as the following sum,
\begin{equation}\label{sum-eps}
   G_s(m)= \sul_{\epsilon_1+\dots +\epsilon_m-m=s}
                           F_m(\{\epsilon_j,\epsilon_j\}).
\end{equation}
One can immediately remark from \eq{EB-2S} that the ground state
density functional $ \mathcal{S}(\{\la\})$ \eq{densityM}  is
common for all the terms in this sum. After symmetrisation over
the variables $\lambda$ corresponding to the integrals of the same
type and extraction of the common denominator
\begin{align}
  \Theta_m^s(\la_1,\dots,\la_m)= &\prod_{k=1}^s\prod_{j=s+1}^m
          \frac 1{\sinh(\la_j-\la_k)}\nonumber\\
   &\times\pl_{m\ge j>k> s}
     \frac{\sinh(\la_j-\la_k)}{\sinh(\la_j-\la_k+\eta)\sinh(\la_j-\la_k-\eta)}
    \nonumber\\
&\times
    \pl_{s\ge j>k\ge 1}
   \frac{\sinh(\la_j-\la_k)}{\sinh(\la_j-\la_k+\eta)\sinh(\la_j-\la_k-\eta)},
\end{align}
we obtain the following representation:
\be{facto} G_s(m)=  \frac{1}{s!(m-s)!}\int_{\cal C}d\la_1\dots
       \int_{\cal C} d\la_m \,\, \Theta_m^s(\la_1,\dots,\la_m)\cdot
                                 \mathcal{G}_s(m,\{\la\}|\{\xi\})\cdot
                                 \mathcal{S}(\{\la_j\}).
\ee

The function $\mathcal{G}_s(m,\{\la\}|\{\xi\})$ in \eq{facto} is a
rather complicated sum over permutations which corresponds to the
sum \eq{sum-eps} over all possible configurations of the algebraic
part in the expression \eq{EB-2S} of the elementary blocks. It is
possible to express it in a simpler form if we notice that it
satisfies the four following important properties:
\begin{enumerate}
\item The function $\mathcal{G}_s(m,\{\la\}|\{\xi\})$  is symmetric under the
permutations  of the variables $\xi_1$, $\xi_2,\dots ,\xi_m$.
\item The function $e^{(m-1)\la_j}\,\mathcal{G}_s(m,\{\la\}|\{\xi\})$ is a
polynomial function of $e^{2\la_j}$ of degree $m-1$.
\item For $m=1$,
\begin{equation}
  \mathcal{G}_0(1,\la_1|\xi_1)=\mathcal{G}_1(1,\la_1|\xi_1)=1.
\end{equation}
\item The function $\mathcal{G}_s(m,\{\la\}|\{\xi\})$ satisfies the
following recursion relations,
\begin{align}
 &\mathcal{G}_s (m,\{\la\}|\{\xi\})\pour{\la_j=\xi_k}
      =\pl_{\substack{a=1\\a\neq k}}^m
       \sinh(\la_j-\xi_a+\eta)\pl_{a\neq j}
       \sinh(\la_a-\xi_k+\eta)\nonumber\\
&\ \,\times
  \mathcal{G}_{s}(m-1,\la_1,\dots,\la_{j-1},\la_{j+1},\dots,\la_m|
                             \xi_1,\dots,\xi_{k-1},\xi_{k+1},\dots\xi_m),
 \qquad
 j\le s, \label{recurs1}
    \\
 &\mathcal{G}_s(m,\{\la\}|\{\xi\})\pour{\la_j=\xi_k}
      \vphantom{\left.\pl_{j=1}\right|_{\la_j=\xi_k}}
     =\pl_{\substack{a=1\\a\neq k}}^m
      \sinh(\la_j-\xi_a+\eta)\pl_{a\neq j}
      \sinh(\la_a-\xi_k+\eta)\nonumber\\
&\ \,\times
  \mathcal{G}_{s-1}(m-1,\la_1,\dots,\la_{j-1},\la_{j+1},\dots,\la_m|
                               \xi_1,\dots,\xi_{k-1},\xi_{k+1},\dots\xi_m),
\quad
 j>s.\label{recurs2}
\end{align}
\end{enumerate}
These properties can be easily proved using the definition of
$\mathcal{G}_s(m,\{\la\}|\{\xi\})$. They define this function in a
unique way as for any $m$ they define a polynomial of degree $m-1$
in $m$ points. Recursion relations of the same kind as
\eq{recurs1}-\eq{recurs2} were obtained for the first time by
Korepin in \cite{Kor82} for the partition function of the
six-vertex model with domain wall boundary conditions, and the
corresponding unique solution was found by Izergin in
\cite{Ize87}. The conditions 1-4 are very similar to the
conditions that characterise the partition function except that
they contain one more parameter $s$. However,  the expression for
the partition function obtained by Izergin satisfies these
relations for any $s$. As the solution  of the recursion relation
is unique, we can conclude that the function
$\mathcal{G}_s(m,\{\la\}|\{\xi\})$ is proportional to the
partition function $Z_m(\{\la\},\{\xi\})$ and does not depend on
$s$. More precisely,
\begin{equation}
  \mathcal{G}_s(m,\{\la\}|\{\xi\})=\frac 1{\sinh^m\eta}\, Z_m(\{\la\},\{\xi\}),
\end{equation}
where the partition function is given by the Izergin formula,
\be{part-fun}
   Z_m(\{\la\},\{\xi\})
  =\frac {\pl_{j=1}^m\pl_{k=1}^m\sinh(\la_j-\xi_k+ \eta)\sinh(\la_j-\xi_k)}
         {\pl_{j>k}^m\sinh(\la_j-\la_k)\sinh(\xi_k-\xi_j)}
   \cdot\det_m\, [ t(\la_j,\xi_k)].
\ee

 We obtain finally the generating function $\bracket{Q^\kappa_{1,m}}$
as a sum of $m+1$ terms, each of them being given as a $m$-fold
multiple integral:
\begin{align}\label{repr2}
   &\bracket{Q^\kappa_{1,m}}=\sul_{s=0}^m \kappa^s G_s(m),\\
   &G_s(m)= \frac{1}{s!(m-s)!\sinh ^m\eta}\int\limits_{\cal C}d^m\la
       \,\,\Theta_m^s(\la_1,\dots,\la_m)\cdot
              Z_m(\{\la\}|\{\xi\})\cdot\mathcal{S}(\{\la\}).
\end{align}
It is interesting to mention that the first and the last terms in
this sum give a representation for the emptiness formation
probability which will be studied in details in the next section.
One of the most interesting property of this representation  is
the presence under the integrals of the expression for the
partition function of the corresponding six-vertex model with
domain wall boundary conditions. This is a new and unexpected
connection of this very important object with the correlation
functions of the $XXZ$ spin chain.

One can note that the two representations \eq{Q-int-rep} and
\eq{repr2} of the generating function $\bracket{Q^\kappa_{1,m}}$
that we have obtained in this section are quite different and
present different advantages: the first terms of \eq{Q-int-rep}
are very simple, but further terms become more and more
complicated, whereas all the terms of \eq{repr2} have more or less
the same structure. One can hope that this last remark may lead to
a common strategy to compute their asymptotics.

\bigskip

Similar expressions can be obtained for the two-point functions.
For example the correlation function $g_{+
-}(m)=\bracket{\sigma_1^+\,\sigma_{m+1}^-}$ can be written as
\begin{equation}
g_{+ -}(m)= \sul_{s=0}^{m-1}  \tilde{g}_{+-}(m,s) ,
\end{equation}
in which the coefficients $\tilde{g}_{+-}(m,s)$ are given as the
following multiple integrals,
\begin{align}
\tilde{g}_{+-}(m,s)=& \frac 1{s!(m-1-s)!\sinh^{m-1}\eta}
\int\limits_{\cal C}\! d\la_2\dots \int\limits_{\cal C}\! d\la_m
\int\limits_{\cal C}\! d\la_+ \int\limits_{\cal C}\! d\la_-
         \nonumber\\
&\qquad\times \left(\pl_{k=2}^{s+1} \frac{\sinh(\la_- -\xi_k+\eta
) \sinh(\la_k-\xi_1+\eta )}{\sinh(\la_- -\la_k+\eta)}\right)
         \nonumber\\
&\qquad\times \left(\pl_{k=s+2}^{m} \frac{\sinh(\la_- -\xi_k+\eta
) \sinh(\la_k-\xi_1)}{\sinh(\la_- -\la_k)} \right)
         \nonumber\displaybreak[0]\\
&\qquad\times \left(\pl_{k=2}^{s+1} \frac{\sinh(\la_+ -\xi_k)
\sinh(\la_k-\xi_{m+1})}{\sinh(\la_+ -\la_k)}\right)
         \nonumber\displaybreak[0]\\
&\qquad\times \left(\pl_{k=s+2}^{m} \frac{\sinh(\la_+ -\xi_k)
\sinh(\la_k-\xi_{m+1}+\eta )}{\sinh(\la_+ -\la_k-\eta)}\right)
         \nonumber\displaybreak[0]\\
&\qquad\times \frac{\sinh(\la_+ -\xi_1) \sinh(\la_- -\xi_1+\eta
)}{\sinh(\la_- -\la_+)}\,\,\cdot \Theta_{m-1}^s(\la_2,\dots,\la_m)
         \nonumber\\
&\qquad\times
Z_{m-1}(\{\la_2,\dots,\la_m\},\{\xi_2,\dots,\xi_m\})\cdot
\mathcal{S}(\{\la_2,\dots,\la_m,\la_+,\la_-\}).\label{resum+-f}
\end{align}
A very similar representation can be also obtained directly for
the two-point function $g_{z z}(m)$.



\section{Towards asymptotic analysis}\label{sec-4}

We have seen in the last section that it was possible to re-sum,
at least partially, the multiple integral representation for the
two-point function given by the sum over elementary blocks. This
provides of course a more compact expression but, above all, an
expression that seems, due to the particular form of the resulting
multiple integrals, more suitable for the study of the asymptotic
behaviour at large distances. In this section, we will see on a
simple example how it is indeed possible to analyse this kind of
integrals. We then discuss the problems that arise when one tries
to extend this study to either representation \eq{Q-int-rep} or
\eq{repr2} of the two-point function.

\subsection{A simple example: the emptiness formation probability}

There exists a particular correlation function for which it is
possible to compute the main asymptotic behaviour: the so-called
emptiness formation probability $\tau(m)$, which measures the
probability of formation of some ferromagnetic sub-chain of length
$m$ in the (anti-ferromagnetic) ground state. It is defined as the
expectation value
\be{EMPtau}
   \tau(m)=\bra{\psi_g}\,\prod_{k=1}^m\frac{1-\sigma_k^z}2 \,\ket{\psi_g}
\ee
on the normalised ground state $\ket{\psi_g}$ of the chain. Hence,
this quantity corresponds to a single elementary block, which
means that, in the framework of Section \ref{sec-2}, it is given
as a (single) multiple integral of the type \eq{resultEB} with $m$
integrations \cite{JimM95L,KorIEU94,KitMT00}:
\begin{multline}\label{EMPbloc}
   \tau(m)=\lim_{\xi_1,\dots\xi_m\to\eta/2} \
    \prod\limits_{a<b}^m \frac{1}{\sinh(\xi_a-\xi_b)}\\
      \times\int\limits_{\cal C} d^m\lambda\
 \frac{\prod\limits_{j=1}^{m}\bigg\{
       \prod\limits_{k=1}^{j-1} \sinh(\lambda_{j}-\xi_k+\eta)
       \prod\limits_{k=j+1}^{m} \sinh(\lambda_{j}-\xi_k)\bigg\}}
      {
       \prod\limits_{a>b}^m \sinh(\lambda_{a}-\lambda_{b}+\eta)}\
 \det_m[\rho(\lambda_j,\xi_k)] .
\end{multline}
Due to its combinatorial simplicity, it has been widely studied
recently (see for example \cite{BooK01, RazS01, ShiTN01,
KitMST02b, BooKNS02, BooKS03}). However, the expression
\eq{EMPbloc} is not convenient for the asymptotic analysis; in
particular it is not symmetric. Its symmetrised version, obtained
in \cite{KitMST02a}, follows directly from the limit
$\kappa\tend\infty$ in representations \eq{Q-int-rep} or
\eq{repr2} of the generating function $\bracket{Q_{1,m}^\kappa}$:
\begin{multline}\label{EFPsym1}
\tau(m)=\lim_{\xi_1,\dots\xi_m\to\eta/2}
  \frac1{m!} \int\limits_{\cal C} d^m\lambda\
  \prod\limits_{a,b=1}^m  \frac{1}{\sinh(\lambda_a-\lambda_b+\eta)}
                 \\
\times
  \prod\limits_{a<b}^m \frac{\sinh(\lambda_a-\lambda_b)}
                            {\sinh(\xi_a-\xi_b)}
   \cdot
   Z_m(\{\lambda\},\{\xi\})\cdot
   \det_m [\rho(\lambda_j,\xi_k)],
\end{multline}
where $  Z_m(\{\lambda\},\{\xi\})$ denotes the partition function
of the six-vertex model with domain wall boundary conditions given
by \eq{part-fun}. From this expression, it is possible to obtain
the asymptotic behaviour of $\tau(m)$ using the saddle-point
method. This was performed for the first time in \cite{KitMST02b}
in the case of free fermions ($\Delta=0$), but the method of
\cite{KitMST02b} can be applied to the general case as well (see
\cite{KitMST02d} for the study in the massless regime). We briefly
recall here the main step of this computation and present the
result in massless and massive regime.

To apply the saddle-point method to  \eq{EFPsym1}, it is
convenient to express the integral in the following form:
\be{EFPsym2} \tau(m)=\int\limits_{\cal D} d^m\lambda\
    G_m(\{\lambda\})\ e^{m^2 S_m(\{\lambda\})},
\ee
with
\begin{align}
  S_m(\{\lambda\}) =& -\frac{1}{m^2}\sul_{a>b}^m
   \log[\sinh(\la_a-\la_b+\eta)\, \sinh(\la_a-\la_b-\eta)]
               \nonumber\\
        &+\frac{1}{m}\sul_{a=1}^m
       \log[\sinh(\la_a+\eta/2)\, \sinh(\la_a-\eta/2)]
               \nonumber\\
        &+\frac{1}{m^2}
         \build\lim_{\xi_1\ldots\xi_m\to\eta/2}^{}
      \log\Big[
        \Big(\frac{-2 i \pi}{\sinh\eta}\Big)^m
        \frac{\big(\det\rho(\la_j,\xi_k)\big)^2}
             {\pl_{a\ne b}\sinh(\xi_a-\xi_b)}
            \Big]
               \label{EFP-S}
\end{align}
and
\be{EPF-G} G_m(\{\la\})= \build\lim_{\xi_1\ldots\xi_m\to\eta/2}^{}
        \frac{ \det_m\big[ \frac{i}{2\pi} t(\la_j,\xi_k) \big] }
     {\det_m\rho(\la_j,\xi_k)}.
\ee
In \eq{EFPsym2}, the integration domain $\mathcal{D}$ is such that
the variable of integration $\la_1,\ldots,\la_m$ are ordered in
the interval $\mathcal{C}=[-\Lambda_h,\Lambda_h]$ (i.e.
$-\Lambda_h<\la_1<\cdots<\la_m<\Lambda_h$ in the massless case,
and $-i\Lambda_h< i\la_1<\cdots<i\la_m< i\Lambda_h$ in the massive
case).

In the case of free fermions ($\Delta=0$), $G_m(\{\la\})\equiv 1$,
and it is easy to see that $S_m$ admits a unique maximum
$S_m(\{\lambda'\})$ for a set of variables
$\{\la'_1,\ldots,\la'_m\}$ satisfying the system of $m$
saddle-point equations:
\be{saddle-eq}
  \partial_{\la_j} S_m(\{\la'\}) =0,\qquad 1\le j\le m.
\ee
In the limit $m\to\infty$, the distribution of these variables
$\lambda'$s at the saddle point can be described by a density
function,
%
\be{saddle-dens}
  \rho_s(\lambda'_j)=\lim_{m\to\infty}
   \frac1{m(\lambda'_{j+1}-\lambda'_j)},
\ee
and one can replace sums over the set $\{\lambda'\}$ by integrals:
\begin{align}
  &\frac{1}{m}\sum_{j=1}^{m} f(\la'_j)
       \xrightarrow[m\to\infty]{}
       \int\limits_{\cal C} f(\la)\rho_s(\la) d\la,
         \label{sumint1}\\
  &\frac{1}{m}\sum_{\substack{j=1\\j\ne k}}^{m}
       \frac{f(\la'_j)}{\la'_j-\la'_k}
     \xrightarrow[m\to\infty]{}
      V.P. \int\limits_{\cal C} \frac{f(\la)}{\la-\la'_k}\rho_s(\la) d\la,
        \label{sumint2}
\end{align}
for any function $f$ integrable on the contour ${\cal C}$. Hence,
the system \eq{saddle-eq} becomes a single integral equation for
the density $\rho_s(\lambda')$, that can be solved explicitely by
Fourier transform. Replacing, at the leading order in $m$, the
expression of this saddle-point density in the integrals that
approximate the sums in \eq{EFP-S}, one obtains that the main
behaviour of the emptiness formation probability at the free
fermion point in a magnetic field $h$ ($|h|<4$) is given by
(see  \cite{KitMST02b} for details)\footnote{%
For $|h|\ge4$ the ground state becomes ferromagnetic and the
emptiness formation probability is equal to $0$ (for $h\ge4$) or
to $1$ (for $h\le-4$).}
\be{asymp-ff}
   \frac{1}{m^2} \log\tau(m) \ \build\sim_{m\to\infty}^{} \
         S^{(0)}=\frac{1}{2} \log\Big(\frac{4-h}{8}\Big).
\ee

The general case is slightly more complicated, but follows the
same procedure. The main problem is that, a priori, we do not know
any asymptotic equivalent of the quantity $G_m({\la})$ when
$m\to\infty$. Nevertheless, in the case of zero magnetic field, it
is still possible to compute the asymptotic behaviour of
\eq{EFPsym2} in the leading order, provided we make the following
hypothesis: we assume that the integrand of \eq{EFPsym2} admits a
maximum for a certain value $\la_1',\dots,\la_m'$ of the
integration variables $\la_1,\dots,\la_m$, that, for large $m$,
the distribution of these parameters $\la_1',\dots,\la_m'$ can be
described by a  density function $\rho_s(\la')$ of the form
\eq{saddle-dens} on the symmetric interval $[-\Lambda,\Lambda]$
(see \eq{inf1}, \eq{inf2}), and that, at the leading order in
$m$, we can replace the sums over the set of parameters $\{\la'\}$
by integrals over this density $\rho_s(\la')$ as in
\eq{sumint1}-\eq{sumint2}.

First, like in the free fermion case, it is easy to determine the
maximum of the function $S_m(\{\la\})$. Indeed, let
$\{\tilde\la\}$ be solution of the system
\be{sadp}
   \partial_{\lambda_j} S_m (\{\tilde\la\})=0,\qquad 1\le j\le m.
\ee
In the limit $m\tend\infty$, if we suppose again that the
parameters $\tilde\la_1,\ldots,\tilde\la_m$ become distributed
according to a certain density $\tilde\rho_s(\la)$ and that sums
over the $\tilde\la_j$ become integrals over this density, the
system \eq{sadp} turns again into a single integral equation for
$\tilde\rho_s$, that can be solved explicitely in the case of zero
magnetic field:
%
%
%
\begin{xalignat}{2}\label{rhotildem}
   \tilde\rho_s(\la) &=\frac{i}{\pi}\sum_{n\in\Zset}
                  \frac{\ch(n\zeta)}{\ch(2n\zeta)}\, e^{-2n\la},
      & &(\text{massive case }\ \Delta>1,\ \zeta=-\eta>0),  \\
   &=\frac{\ch\frac{\pi\la}{2\zeta}}{\zeta\sqrt{2}\ch\frac{\pi\la}{\zeta}},
      & &(\text{massless case }|\Delta|<1,\ \zeta=i\eta>0).\label{rhotildeless}
\end{xalignat}
This gives for the maximum of $S_m(\{\la\})$ when
$m\tend\infty$\footnote{At this main order in $m$, there exists a
unique solution of the integral equation for $\tilde\rho_s$, and
we know it corresponds to a maximum because $S_m(\{\la\})\tend
-\infty$ on the boundary of $\mathcal{D}$.}:
\begin{xalignat}{2}
  \lim_{m\tend\infty} S_m(\{\tilde\la\})
   &=-\frac{\zeta}{2}
    -\sum_{n=1}^\infty \frac{e^{-n\zeta}}{n}
                       \frac{\sh(n\zeta)}{\ch(2n\zeta)},\label{maxS-mass}
      & &(\Delta=\cosh\zeta>1),  \\
   &=\log\frac\pi\zeta+
      \frac{1}{2}\int\limits_{\mathbb{R}-i0}
      \frac{d\omega}{\omega}\frac
        {\sinh\frac{\omega}{2}(\pi-\zeta)\cosh^2\frac{\omega\zeta}{2}}
        {\sinh\frac{\pi\omega}{2}\sinh\frac{\omega\zeta}{2}\cosh\omega\zeta},
      & &(|\Delta=\cos\zeta|<1).\label{maxS-less}
\end{xalignat}

The second step is to show that the factor $G_m(\{\la\})$ gives
always a negligible contribution compared to $S_m(\{\tilde\la\})$
at this order in $m$, at least for any distribution of the
variables $\la_j$ satisfying the previous hypothesis of
regularity. Indeed, we can use the integral equation
\eq{GFinteqinh} satisfied by the inhomogeneous spectral density
for the ground state to express, for any set of variables
$\{\la\}$,  $G_m(\{\la\})$ in the form:
\be{GLieb}
   G_m(\{\la\}) = \lim_{\xi_1,\dots\xi_m\to-\frac{i\zeta}2}
  \frac{{\det}_m \left[\rho(\la_j,\xi_k)
          +\int_{\cal C}K(\la_j-\mu)\rho(\mu,\xi_k)d\mu\right]}
       {{\det}_m [\rho(\la_j,\xi_k)]},
\ee
where the kernel $K$ is given by \eq{GFKkern}. If the distribution
of $\{\la\}$ is regular enough in the interval
$[-\Lambda,\Lambda]$, we can replace, in the limit $m\tend\infty$,
the integral
\be{intKrho}
   \int\limits_{\cal C} K(\la_j-\mu)\rho(\mu,\xi_k)d\mu
\ee
in the determinant by the sum
\be{sumKrho}
  \frac{1}{m}\sum_{l=1}^m K(\la_j-\la_l)
                     \frac{\rho(\la_l,\xi_k)}{\hat\rho_s(\la_l)}
\ee
where the density function $\hat\rho_s(\la)$ describes the
distribution of the $\la_j,\ j=1,\ldots,m$ in the limit
$m\tend\infty$. Therefore, \be{equivG}
  G_m({\la})\build\sim_{m\to\infty}^{}
   \det_m\Big(\delta_{jk}+\frac{K(\la_j-\la_k)}{m\ \hat\rho_s(\la_k)}\Big).
\ee In the massive regime, this is merely the Fredholm determinant
of the integral operator $\hat{I}+\hat{K}$, where $\hat{I}$
denotes the identity operator, and $\hat{K}$ the integral operator
of kernel $K$ \eq{GFKkern}. This determinant is given by the
infinite product of its eigenvalues:
\be{limG}
  \lim_{m\tend\infty}  G_m(\{\la\})=\det(\hat{I}+\hat{K})
                               =2\prod_{n=1}^\infty(1+q^{2n})^2,
  \qquad q=e^{\eta}\quad(\text{massive regime}).
\ee
In the massless regime, the determinant \eq{equivG} and its
inverse can be bounded via Hadamard inequality. Thus, in both
regime, we can show that
\be{negG}
   \lim_{m\tend\infty}\frac{1}{m^2}\log G_m(\{\la\}) =0
\ee
for any distribution of $\{\la\}$ with good properties of
regularity, in particular for the saddle point. This means that,
at the main order in $m$, the factor  $G_m(\{\la\})$ does not
contribute to the value of the maximum of the integrand, and that
the latter is indeed given by the maximum
\eq{maxS-mass}-\eq{maxS-less} of $S_m(\{\la\})$, $\tilde\rho_s$
being identified with the saddle-point density $\rho_s$.

Finally we obtain the following result concerning the asymptotic
behaviour of $\tau(m)$ for $m\to\infty$ (see \cite{KitMST02d} for
the massless case):
\begin{alignat}{2}
  S^{(0)}(\Delta) &=\lim_{m\to\infty}\frac{\log\tau(m)}{m^2}, & & \\
    &=-\frac{\zeta}{2}
    -\sum_{n=1}^\infty \frac{e^{-n\zeta}}{n}
                       \frac{\sh(n\zeta)}{\ch(2n\zeta)},\label{asympt-mass}
      &\quad &(\Delta=\cosh\zeta>1),  \\
    &=\log\frac\pi\zeta+
      \frac{1}{2}\int\limits_{\mathbb{R}-i0}
      \frac{d\omega}{\omega}\frac
         {\sinh\frac{\omega}{2}(\pi-\zeta)\cosh^2\frac{\omega\zeta}{2}}
         {\sinh\frac{\pi\omega}{2}\sinh\frac{\omega\zeta}{2}\cosh\omega\zeta},
      &\quad &(-1<\Delta=\cos\zeta<1).\label{asympt-less}
\end{alignat}
Note that this coincides with the exact known results obtained in
\cite{ItsIKS93,ShiTN01,KitMST02b} at the free fermion point and in
\cite{RazS01,KitMST02c} at $\Delta=1/2$, and is in agreement with
the expected value in the Ising limit:
\begin{alignat}{2}
  &S^{(0)}(\Delta=0)= -\frac{1}{2}\log2
          & \qquad  &(\text{Free fermion case}),\\
  &S^{(0)}(\Delta=\frac{1}{2})= \frac{3}{2}\log3-3\log2,\\
  &S^{(0)}(\Delta) \xrightarrow[\Delta\to \infty]{}  -\infty
          & \qquad  &(\text{Ising case}).
\end{alignat}
Moreover, we can apply the same saddle-point procedure directly at
the $XXX$ point $\Delta=1$ and check that
\begin{align}
  S^{(0)}(\Delta=1)
    &= S^{(0)}(\Delta\tend 1^+) = S^{(0)}(\Delta\tend 1^-)\nonumber\\
    &= \log \biggl(
         \frac{\Gamma\left(\frac{3}{4}\right)\Gamma\left(\frac{1}{2}\right)}
              {\Gamma\left(\frac{1}{4}\right)}\biggr)\approx\log(0.5991),
\end{align}
which is in good agreement with the numerical result
$\log(0.598)$, obtained in \cite{BooKNS02}.

In the massless regime, the leading asymptotic behaviour
\eq{asympt-less}, was conjectured independently in
\cite{KorLNS03}. In that article was also conjectured the first
(power-law) sub-leading correction in the form:
\begin{equation}\label{conj1}
  \tau(m) \build\sim_{m\to\infty}^{} A\, m^{-\gamma} e^{-m^2 S_0},\qquad
    (-1<\Delta=\cos\zeta<1),
\end{equation}
with
\be{conj2}
   \gamma=\frac{1}{12}+\Big(\frac\zeta\pi\Big)^2 \frac1{3(1-\zeta/\pi)},
\ee
which is in agreement with the exact results at $\Delta=0$ and
$\Delta=1/2$ (see \cite{ShiTN01}, \cite{KitMST02c}). It would be
interesting to check this latter conjecture by analysing
corrections to the saddle-point method we have presented here.
%

\subsection{The two-point functions: attempts and problems}

The long-distance asymptotics of physical correlation functions,
such as the two-point functions, have attracted long-standing
interest. In the case of the $XXZ$ model, some predictions were
made already a long time ago.

In the massive regime ($\Delta>1$), spin-spin correlation
functions are expected to decay exponentially with the distance
and the exact value of the correlation length was proposed in
\cite{JohKM73}. For the $XXZ$ chain in the massless regime
($-1<\Delta\le 1$), zero temperature is a critical point and the
correlation length becomes infinite in units of the lattice
spacing. The leading long-distance effects can be predicted by
conformal field theory and the correlation functions are expected
to decay as a power of the distance. In particular, one expects
that, at the leading order,
\begin{align}
  &\bracket{\sg_j^x\,\sg_{j+n}^x}=(-1)^n\frac{A}{n^{\pi-\zeta}}+\cdots,
              \label{asympt-x}\\
  &\bracket{\sg_j^z\,\sg_{j+n}^z}= -\frac{1}{\pi(\pi-\zeta)}\frac{1}{n^2}
              + (-1)^n\frac{A_z}{n^\frac{\pi}{\pi-\zeta}}+\cdots.
               \label{asympt-z}
\end{align}
A conjecture for the non-universal correlation amplitudes $A$ and
$A_z$ can be found in \cite{LukZ97,Luk99,LukT03}. The exact value
of the critical exponents in \eq{asympt-x}-\eq{asympt-z} was
proposed for the first time in \cite{LutP75}.

However, there does not exist at the moment any direct derivation
of these predictions from the exact expressions of the correlation
functions on the lattice. In the last subsection we have shown how
to determine, at least in the main order, the asymptotic behaviour
of the emptiness formation probability using the saddle-point
method. We could expect to be able to apply the same technique to
the new multiple integral representation of the two-point function
obtained in Section~\ref{sec-3}.

In particular, one can notice immediately that each term of the
representation \eq{repr2} of the generating functional
$\bracket{Q_{1,m}^\kappa}$ has a structure very similar to
\eq{EFPsym1}. Indeed, it is possible to apply to the whole sum a
slight modification of the saddle-point technique presented here.
It shows that, as it should be, there is no contribution of order
$\exp(\alpha m^2)$ when $m\to\infty$.
 However, to obtain the precise asymptotic behaviour of the
two-point function, one should be able to analyse sub-leading
corrections to this saddle-point method, which is technically
quite difficult. It is not obvious in particular from these
expressions that, in the massless regime, the leading asymptotic
behaviour of the two-point function is only of power-law order.

If one considers instead \eq{Q-int-rep}, one can try to make a
similar analysis for each term of the sum. It can be easily proved
that, in the massless regime, the first terms of \eq{Q-int-rep}
decrease as powers of the distance. However, it is neither
difficult to see that the next terms of the series are not
negligible with respect to the first ones, which means that one
should analyse the whole sum to obtain the correct power law
asymptotic behaviour.



\section{Complete re-summation for the finite chain}\label{sec-5}

We have  seen in the last section that, although the partial
re-summations presented in Section~\ref{sec-3} contain integrals
that can be analysed in the main order via the saddle-point
method, the asymptotic analysis of the sum itself is much more
tricky. It is due to the fact that we have to take into account
sub-leading corrections to the saddle-point to be able to obtain
merely the main order asymptotic of either representation
\eq{Q-int-rep} or \eq{repr2}. From this point of view, it could be
more convenient to deal only with one single (multiple) integral
instead of a sum, as in the case of the emptiness formation
probability.

It is actually possible to re-sum completely representation
\eq{Q-interm} to obtain, at the finite chain level, what we call
the {\em master equation} for the two-point function. We will see
in particular that this master equation sheds a new light on the
previous algebraic re-summation of Section~\ref{sec-31} by
connecting it to the expansion of the two-point function in terms
of the form factors of the local spin operators.

\subsection{Multiple action of transfer matrices: a complete re-summed
formula}\label{sec-51}

To perform the re-summation  of the two-point function, we have
used in Section~\ref{sec-31} the explicit formula \eq{APcompl-act}
for the multiple action of $\prod_{a=1}^m{\cal T}_\kappa(x_a)$ for
an arbitrary set of complex numbers $\{x\}$ on an arbitrary state
$\bra{0}\prod_{j=1}^N C(\mu_j)$ \cite{KitMST02a}. Introducing
auxiliary integrals as in \eq{GFcontint}, one can actually re-sum
completely the sum over the partitions of the sets $\{\mu\}$ and
$\{x\}$ in \eq{APcompl-act} and express this action in the form of
a single multiple integral:

\begin{prop}\label{lemme-actTM}
Let $\kappa$, $x_1,\ldots,x_m$ and $\mu_1,\ldots,\mu_N$ be generic
parameters. Then the action of $\prod_{a=1}^m{\cal T}_\kappa(x_a)$
on a state of the form $\bra{0}\prod_{j=1}^NC(\mu_j)$ can be
formally written as
 \begin{multline}\label{actTM}
 \bra{0}\prod_{j=1}^N C(\mu_j)\prod_{a=1}^m{\cal T}_\kappa(x_a)
    = \frac{1}{N!}
    \oint\limits_{\Gamma\{x\} \cup \Gamma\{\mu\}}
    \prod_{j=1}^N\frac{dz_j}{2\pi i}\cdot
    \prod_{a=1}^{m} \tau_\kappa (x_a| \{z\} ) \cdot
    \prod_{a=1}^{N} \frac{1}{ {\cal Y}_\kappa (z_a| \{z\} )}
\\
 \times
    \prod_{\substack{j,k=1\\ j<k}}^{N}
  \frac{\sh(z_j-z_k)}{\sh(\mu_j-\mu_k)}\cdot
 \det_N \Omega_\kappa ( \{z\} , \{\mu\} | \{z\}  )\cdot
  \bra{0}\prod_{j=1}^N C(z_j),
\end{multline}
where the integration contour $\Gamma\{x\} \cup \Gamma\{\mu\}$
surrounds the points\footnote{%
More precisely, for a set of complex variables
$\{\nu_1,\ldots,\nu_l\}$,  the notation $\Gamma\{\nu\}$  should be
understood in the following way: $\Gamma\{\nu\}$ is the boundary
of a set of poly-disks ${\cal D}_a(r)$  in $\mathbb{C}^N$, i.e.
$\Gamma\{\nu\}={\cup}_{a=1}^l\bar{\cal D}_a(r)$ with $\bar{\cal
D}_a(r)=\{z\in\mathbb{C}^N: |z_k-\nu_a|=r,\quad k=1,\dots,N\}$. }
$x_1,\dots,x_m$  and $\mu_1,\dots,\mu_N$ and does not contain any
other pole of the integrand.
\end{prop}

{\sl Proof}.  The right-hand side in \eq{actTM} can be computed by
the sum over the residues at the points $x_1,\dots,x_m$,
$\mu_1,\dots,\mu_N$. More precisely, it is given by
\be{residus}
  (2i\pi)^N \sum_{n=0}^{\min(m,N)}
  \sum_{\substack{
         \{\mu\}=\{\mu_{\alpha_+}\}\cup\{\mu_{\alpha_-}\}\\
         \{x\}=\{x_{\gamma_+}\}\cup\{x_{\gamma_-}\}\\
         |\alpha_+|=|\gamma_+|=n}}
  \Res_{\{z\}=\{x_{\gamma_+}\}\cup\{\mu_{\alpha_-}\}}(\text{Integrand}),
\ee
and one can easily check that the residue of the integrand for
$\{z\}=\{x_{\gamma_+}\}\cup\{\mu_{\alpha_-}\}$ is equal to
$R_n^\kappa(\{x_{\gamma_+}\},\{x_{\gamma_-}\},
\{\mu_{\alpha_+}\},\{\mu_{\alpha_-}\})$. \qed

To apply this formula to the effective computation of the
two-point functions, one should give specific values to the
parameters $x_1,\ldots,x_m$, $\mu_1,\dots,\mu_N$: they will
correspond in this context either to admissible solutions of the
Bethe equation at $\kappa=1$ (parametrising the eigenstate in
which we compute expectation values) or to inhomogeneity
parameters (set to $\eta/2$ in the homogeneous case) arising from
the reconstruction of local spin operators. To ensure the
existence of the corresponding contour $\Gamma\{x\} \cup
\Gamma\{\mu\}$ for these particular values, one has to prove that
there exists a surrounding of
$[\{x_1,\ldots,x_m\}\cup\{\mu_1,\ldots,\mu_N\}]^N$ which does not
contain any new singularity of the integrand. Since $C(z)$ is a
polynomial in $z$ in the normalisation \eq{op-L}, such
singularities could occur only due to the product of the twisted
Bethe equations ${\cal  Y}_\kappa (z_a| \{z\} )$ in the
denominator. We will see in the following that these poles play a
central role in the context of master equation.

Solutions of the system of Bethe equations \eq{TTMBE_Y} were
studied in \cite{TarV96} (see also Appendix~A of \cite{KitMST04b}
for the homogeneous case). Following the arguments of
\cite{KitMST04a}, one can easily prove that unadmissible and
diagonal solutions do not correspond to singularities of the
integrand, for either $\det\Omega_\kappa$ or $  \langle 0
|\prod_{j=1}^N C(z_j)$ vanishes in this case.   It was moreover
proven in \cite{KitMST04a,KitMST04b} that, for $|\kappa| $ small
enough, all admissible off-diagonal solutions are in the
vicinities of the shifted inhomogeneity parameters $\xi_j-\eta$
(or of $-\eta/2$ in the homogeneous case) but are separated from
these points as soon as $\kappa\ne 0$. For $|\kappa|$ small
enough, they are also separated from any admissible off-diagonal
solution $\la_1,\ldots,\la_N$ of the system of untwisted Bethe
equations. Thus, just like in \cite{KitMST04a,KitMST04b}, we can
formulate the following lemma:
\begin{lemma}\label{main-lem}
Let $\{\lambda\}$ be an admissible off-diagonal solution of the
system of untwisted Bethe equations, and $\{\xi\}$ be a set of
inhomogeneity parameters. There exists $\kappa_0>0$ such that, for
$|\kappa|<\kappa_0$, one can define a closed contour
$\Gamma\{\lambda\}\cup\Gamma\{\xi\}$ which satisfies the following
properties:

1) it surrounds the points $\{\lambda\}$ and $\{\xi\}$, while all
admissible off-diagonal solutions of the system  \eq{TTMBE_Y} are
outside of this contour;

2) the only poles which are inside and provide non-vanishing
contributions to the integral \eq{actTM} are $z_j=\lambda_k$ and
$z_j=\xi_k$;

3) the only poles which are outside (within a set of periodic
strips) and provide non-vanishing contributions to the integrand
of \eq{actTM} are the admissible off-diagonal solutions of the
system of twisted Bethe equations \eq{TTMBE_Y}.
\end{lemma}

\begin{rem}
  Lemma \ref{main-lem} holds also in the homogeneous limit $\xi_j=\eta/2$.
\end{rem}

%
%
%
%
%
%
\subsection{Master equation for the two-point function}\label{sec-52}

Proposition \ref{lemme-actTM} associated to Lemma \ref{main-lem}
enables us to obtain representations for the two-point functions
of the finite chain as a single-multiple integral
\cite{KitMST04a}. Let us consider, for a given eigenstate
$\ket{\psi(\{\la\})}$ of the untwisted transfer matrix, the
expectation values
\begin{align}
  &\bracket{\sigma_1^z\, \sigma_{m+1}^z}=\bracket{\sigma_1^z}
    +\bracket{ \sigma_{m+1}^z}-1
    +2
      \frac{\partial^2}{\partial\kappa^2}
     \bracket{ Q_{1,m+1}^\kappa-Q_{1,m}^\kappa-
               Q_{2,m+1}^\kappa+ Q_{2,m}^\kappa }
         \pour{\kappa=1},
     \label{szsz}\\
  &\bracket{\sigma_1^\alpha\, \sigma_{m+1}^\beta}=\lim_{\kappa\to 1}
    \bracket{\sigma_{1,\kappa}^\alpha\, \sigma_{\kappa, m+1}^\beta},
  \qquad (\alpha,\beta)=(-,+),(+,-).
     \label{smsp}
\end{align}
For convenience, we have defined here the following
$\kappa$-deformed spin operators:
\begin{equation}\label{deformed-spin}
  \sigma_{j,\kappa}^\alpha =\sigma_j^\alpha \cdot
      \prod_{a=1}^{j}{\cal T}(\xi_a)\cdot
      \prod_{a=1}^{j}{\cal T}_\kappa^{-1}(\xi_a),\qquad
  \sigma_{\kappa,j}^\alpha =\prod_{a=1}^{j-1}{\cal T}_\kappa(\xi_a)\cdot
         \prod_{a=1}^{j-1}{\cal T}^{-1}(\xi_a)\cdot
         \sigma_j^\alpha,
\end{equation}
and the operator $Q_{j,k}^\kappa$ is given by \eq{GFdefQ}.
%
%
Using the solution of the quantum inverse scattering problem
\eq{FCtab} and acting with the resulting operators on
$\bra{\psi(\la)}$, notably by means of \eq{actTM}, we obtain the
following result:

\begin{thm}\label{M-thm}
  Let $\{\lambda\}$ be an
  admissible off-diagonal solution of the system of untwisted Bethe
  equations, and let us consider the corresponding expectations values
  \eq{szsz}-\eq{smsp} in the inhomogeneous finite $XXZ$ chain.
  Then there exists $\kappa_0>0$ such that, for $|\kappa|<\kappa_0$,
  the following representations hold:
\begin{multline}\label{master1}
 \bracket{Q_{1,m}^\kappa}=
 \frac{1}{N!}
 \oint\limits_{\Gamma\{\xi\} \cup \Gamma\{\lambda\}}
\prod_{j=1}^N\frac{dz_j}{2\pi i}\cdot
 \prod_{a=1}^{m}\frac{ \tau_\kappa (\xi_a| \{z\} ) }
                     { \tau (\xi_a| \{\lambda\} ) }\cdot
\prod_{a=1}^{N} \frac{1}
                      { {\cal Y}_\kappa (z_a| \{z\} )}
\\
 \times
 \det_N \Omega_\kappa ( \{z\} , \{\lambda\} | \{z\}  )
    \cdot
    \frac{ \det_N \Omega (\{\lambda\} , \{z\}  |\{\lambda\} )}
         { \det_N \Omega (\{\lambda\} , \{\lambda\} | \{\lambda\})},
\end{multline}
%
%
\begin{multline}\label{master2}
  \bracket{\sigma_{1,\kappa}^-\, \sigma_{\kappa, m+1}^+}
   =
   \frac{1}{(N-1)!} \oint\limits_{\Gamma\{\xi\} \cup \Gamma\{\lambda\}}
    \prod_{j=1}^{N-1}\frac{dz_j}{2\pi i}\cdot
    \frac{\pl_{a=2}^{m} \tau_\kappa (\xi_a| \{z\} )}
         {\pl_{a=1}^{m+1} \tau (\xi_a| \{\la\} )} \cdot
    \prod_{a=1}^{N-1} \frac{1}{ {\cal Y}_\kappa (z_a| \{z\} )}
\\
 \times
   \pl_{j=1}^{N-1}\frac{1}{\sh(z_j-\xi_{m+1})}\cdot
   \pl_{j=1}^{N}\frac{1}{\sh(\la_j-\xi_{1})}\cdot
 \frac{\det_N \Omega ( \{\la\} , \{z\}\cup\{\xi_{m+1}\}|\{\lambda\})}
      {\det_N \Omega ( \{\la\} , \{\la\}|\{\lambda\})}
\\
 \times
 \Bigg\{ \sul_{\substack{\alpha,\beta=1\\ \alpha<\beta}}^{N+1}
    (-1)^{\alpha+\beta}\,
     \bigg\{
     \frac{a(\tilde{\la}_\alpha)\, d(\tilde{\la}_\beta)}
          {\sh(\tilde{\la}_\alpha-\tilde{\la}_\beta-\eta)}\cdot
   \pl_{k=1}^{N} \big[ \sinh(\la_k-\tilde{\la}_\alpha+\eta)
                         \sinh(\la_k-\tilde{\la}_\beta-\eta)\big]
\\
  \phantom{\Bigg\{ \sul_{\substack{\alpha,\beta=1\\ \alpha<\beta}}^{N+1}
           (-1)^{\alpha+\beta}\,
           \bigg\{ }
  -\frac{a(\tilde{\la}_\beta)\, d(\tilde{\la}_\alpha)}
        {\sh(\tilde{\la}_\beta-\tilde{\la}_\alpha-\eta)}\cdot
   \pl_{k=1}^{N} \big[ \sinh(\la_k-\tilde{\la}_\beta+\eta)
                         \sinh(\la_k-\tilde{\la}_\alpha-\eta)\big]\bigg\}
\\
 \times
 \det_{N-1} \Omega_\kappa ( \{z\} ,
     \{\tilde{\la}\}\backslash\{\tilde{\la}_\alpha,\tilde{\la}_\beta\}|\{z\} )
            \Bigg\}.
\end{multline}
\begin{multline}\label{master3}
  \bracket{\sigma_{1,\kappa}^+\, \sigma_{\kappa, m+1}^-}
   =
   \frac{1}{(N+1)!}
        \oint\limits_{\Gamma\{\xi\} \cup \Gamma\{\lambda\}}
    \prod_{j=1}^{N+1}\frac{dz_j}{2\pi i}\cdot
    \frac{\pl_{a=2}^{m} \tau_\kappa (\xi_a| \{z\} )}
         {\pl_{a=1}^{m+1} \tau (\xi_a| \{\la\} )} \cdot
    \prod_{a=1}^{N+1} \frac{1}{ {\cal Y}_\kappa (z_a| \{z\} )}
\\
 \times
   \pl_{j=1}^{N+1}\frac{1}{\sh(z_j-\xi_{m+1})}\cdot
   \pl_{j=1}^{N}\frac{1}{\sh(\la_j-\xi_{1})}\cdot
     \frac{\det_{N+1} \Omega_\kappa ( \{z\} , \{\la\}\cup\{\xi_{1}\}|\{z\})}
      {\det_N \Omega ( \{\la\} , \{\la\}|\{\lambda\})}
\\
 \times
 \Bigg\{ \sul_{\substack{\alpha,\beta=1\\ \alpha<\beta}}^{N+2}
    (-1)^{\alpha+\beta}\,
     \bigg\{
     \frac{a(\tilde{z}_\alpha)\, d(\tilde{z}_\beta)}
          {\sh(\tilde{z}_\alpha-\tilde{z}_\beta-\eta)}\cdot
   \pl_{k=1}^{N+1} \big[ \sinh(z_k-\tilde{z}_\alpha+\eta)
                         \sinh(z_k-\tilde{z}_\beta-\eta)\big]
\\
  \phantom{\Bigg\{ \sul_{\substack{\alpha,\beta=1\\ \alpha<\beta}}^{N+2}
           (-1)^{\alpha+\beta}\,
           \bigg\{ }
  -\frac{a(\tilde{z}_\beta)\, d(\tilde{z}_\alpha)}
        {\sh(\tilde{z}_\beta-\tilde{z}_\alpha-\eta)}\cdot
   \pl_{k=1}^{N+1} \big[ \sinh(z_k-\tilde{z}_\beta+\eta)
                         \sinh(z_k-\tilde{z}_\alpha-\eta)\big]\bigg\}
\\
 \times
 \det_{N} \Omega ( \{\la\} ,
     \{\tilde{z}\}\backslash\{\tilde{z}_\alpha,\tilde{z}_\beta\} | \{\la\}  )
   \Bigg\},
\end{multline}
where we have set
$(\tilde{\la}_1,\ldots,\tilde{\la}_{N+1})=(\la_1,\ldots,\la_N,\xi_1)$
and
$(\tilde{z}_1,\ldots,\tilde{z}_{N+2})=(z_1,\ldots,z_{N+1},\xi_{m+1})$
in \eq{master2} and in \eq{master3} respectively. The integration
contours in \eq{master1}-\eq{master3} are such that the only
singularities of the integrand  which contribute to the integral
are the points $\xi_1,\ldots,\xi_m$ and $\lambda_1\ldots,\la_N$.
\end{thm}

In the following, this kind of representations will be called {\em
master
  equation}.


\begin{rem}\label{act-left}
Expressions \eq{master2} and \eq{master3} are obtained by acting
with all the operators on the left, i.e. on the dual Bethe state
$\bra{\psi(\{\la\})}$. It is also possible to act on the right. In
that case, we obtain for $\bracket{\sigma_{1,\kappa}^+\,
\sigma_{\kappa, m+1}^-}$ a representation that is similar to
\eq{master2}, and for $\bracket{\sigma_{1,\kappa}^-\,
\sigma_{\kappa, m+1}^+}$ a representation similar to \eq{master3}.
\end{rem}

\begin{rem}\label{rem-pol}
Note that the $\kappa$-deformed two-point correlation functions
$\bracket{\sigma_{1,\kappa}^\alpha\, \sigma_{\kappa,m+1}^\beta}$,
as well as the generating functional  $\bracket{Q_{1,m}^\kappa}$,
are polynomials in $\kappa$. They are therefore completely
determined by the multiple integral expressions
\eq{master1}-\eq{master3} which are valid at least in a vicinity
of $\kappa=0$. The limit $\kappa\to 1$ can be reached through
analytic continuation.
\end{rem}

\begin{rem}\label{rem-hom}
All these results were formulated in the inhomogeneous case, but
hold also in the homogeneous limit. We can of course choose
$\ket{\psi(\{\la\})}$ to be the ground state of the Hamiltonian of
the $XXZ$ chain.
\end{rem}

\begin{rem}\label{rem-or-series}
From the master equation \eq{master1}, it is easy to come back to
the original series obtained in Section \ref{sec-31}, by
decomposing the multiple integral as
\be{dec-integr}
   \oint\limits_{\Gamma\{\xi\} \cup \Gamma\{\lambda\}}
     \prod_{j=1}^N\,dz_j=
   \sum_{n=0}^N C_N^{\,n}
     \oint\limits_{\Gamma\{\xi\}}\prod_{j=1}^n\,dz_j
     \oint\limits_{\Gamma\{\lambda\}}\prod_{j=1}^{N-n}\,dz_j.
\ee
Note at this stage that, since the number of poles surrounded by
${\Gamma\{\xi\}}$ is $m$ and since the integrand vanishes as soon
as $z_j=z_k$, the sum in \eq{dec-integr} is actually restricted to
$n\le m$. The evaluation of the $N-n$ integrals over the contour
$\Gamma\{\la\}$ as a sum over the residues leads to a sum over the
partitions of the set $\{\la\}$ into two subset
$\{\la_{\alpha_-}\}$ and $\{\la_{\alpha_+}\}$ of cardinal $N-n$
and $n$ respectively. Since the remaining integrals are taken over
the contour $\Gamma\{\xi\}$ that surrounds only the poles at
$z_j=\xi_k$, one can set $d(z_j)=0$ directly in the integrand.
This gives a partial resummation, in which we can use the Bethe
equation for $\{\la\}$ and take the thermodynamic limit as in
Section \ref{sec-31}.

A similar technique can be applied to \eq{master2} or
\eq{master3}.
\end{rem}

\subsection{Form factor expansion}\label{sec-53}

It is also possible, instead of computing the integrals
\eq{master1}-\eq{master3} by the sum over the residues at the
poles inside the contour $\Gamma\{\xi\} \cup \Gamma\{\lambda\}$,
to evaluate them using the poles {\em outside} this contour. Due
to Lemma~\ref{main-lem}, we know that the only poles that can
contribute are the admissible off-diagonal solutions of the system
of twisted Bethe equations \eq{TTMBE_Y}. This leads directly to
the expansion of the correlation functions \eq{szsz}-\eq{smsp}
over form factors.

Let us consider for example the two-point correlation function
$\bracket{\sigma_1^-\, \sigma_{m+1}^+}$. Evaluating \eq{master2}
by the residues at the poles given by the admissible off-diagonal
solutions $\{\mu_1,\ldots,\mu_{N-1}\}$ of the $\kappa$-twisted
Bethe equations outside the integration contour, one obtains
\begin{multline}\label{ev-mu}
  \bracket{\sigma_{1,\kappa}^-\, \sigma_{\kappa, m+1}^+}
   =  \sum_{\{\mu\} }
    \frac{\pl_{a=2}^{m} \tau_\kappa (\xi_a| \{\mu\} )}
         {\pl_{a=1}^{m+1} \tau (\xi_a| \{\la\} )}
\\
 \times
   \pl_{j=1}^{N-1}\frac{1}{\sh(\mu_j-\xi_{m+1})}\cdot
   \pl_{j=1}^{N}\frac{1}{\sh(\la_j-\xi_{1})}\cdot
 \frac{\det_N \Omega ( \{\la\} , \{\mu\}\cup\{\xi_{m+1}\} | \{\lambda\})}
      {\det_N \Omega ( \{\la\} , \{\la\} | \{\lambda\})}
\\
 \times
 \Bigg\{ \sul_{\substack{\alpha,\beta=1\\ \alpha<\beta}}^{N+1}
    (-1)^{\alpha+\beta}\,
     \bigg\{
     \frac{a(\tilde{\la}_\alpha)\, d(\tilde{\la}_\beta)}
          {\sh(\tilde{\la}_\alpha-\tilde{\la}_\beta-\eta)}\cdot
   \pl_{k=1}^{N} \big[ \sinh(\la_k-\tilde{\la}_\alpha+\eta)
                         \sinh(\la_k-\tilde{\la}_\beta-\eta)\big]
\\
  \phantom{\Bigg\{ \sul_{\substack{\alpha,\beta=1\\ \alpha<\beta}}^{N+1}
           (-1)^{\alpha+\beta}\,
           \bigg\{ }
  -\frac{a(\tilde{\la}_\beta)\, d(\tilde{\la}_\alpha)}
        {\sh(\tilde{\la}_\beta-\tilde{\la}_\alpha-\eta)}\cdot
   \pl_{k=1}^{N} \big[ \sinh(\la_k-\tilde{\la}_\beta+\eta)
                         \sinh(\la_k-\tilde{\la}_\alpha-\eta)\big]\bigg\}
\\
 \times
 \det_{N-1} \Omega_\kappa ( \{\mu\} ,
    \{\tilde{\la}\}\backslash\{\tilde{\la}_\alpha,\tilde{\la}_\beta\}|\{\mu\} )
            \Bigg\}.
\end{multline}
with
$(\tilde{\la}_1,\ldots,\tilde{\la}_{N+1})=(\la_1,\ldots,\la_N,\xi_1)$.
In this expression, we can identify the following matrix elements
of $\sigma_{1,\kappa}^-$ and $\sigma_{\kappa,{m+1}}^+$ between the
Bethe state $\ket{\psi(\{\la_j\}_{1\le j \le N})}$ and the
$\kappa$-twisted Bethe state $\ket{\psi_\kappa (\{\mu_j\}_{1\le
j\le N-1})}$:
\begin{align}
 &\bra{\psi_\kappa(\{\mu\})} \, \sigma_{\kappa,{m+1}}^+ \, \ket{\psi(\{\la\})}
  = \frac{\pl_{a=1}^{m}\tau(\xi_a|\{\mu\})}
                                 {\pl_{a=1}^{m+1}\tau(\xi_a|\{\la\})}\,
    \frac{\pl_{j=1}^{N} d(\la_j)}
         {\pl_{\substack{j,k=1\\j<k}}^N \sh(\la_j-\la_k)
          \pl_{\substack{j,k=1\\j<k}}^{N-1} \sh(\mu_k-\mu_j)}
  \nonumber\\
  &\hspace{4.6cm}\times
    \pl_{j=1}^{N-1} \frac{1}{\sh(\xi_{m+1}-\mu_j)}\,
    \det_{N} \Omega(\{\la\},\{\mu\}\cup\{\xi_{m+1}\} | \{\la\}),
           \label{ffp}\displaybreak[0]\\
 &\bra{\psi(\{\la\})} \, \sigma_{1,\kappa}^- \, \ket{\psi_\kappa(\{\mu\})}
    = \frac{1}
         {\tau_\kappa(\xi_1|\{\mu\})}\,
    \frac{\pl_{j=1}^{N-1} d(\mu_j)}
         {\pl_{\substack{j,k=1\\j<k}}^{N-1} \sh(\mu_k-\mu_j)
          \pl_{\substack{j,k=1\\j<k}}^{N} \sh(\la_j-\la_k)
          }
            \nonumber\\
 &\hspace{2.5cm}\times
          \pl_{j=1}^{N} \frac{1}{\sh(\la_j-\xi_1)}
   \Bigg\{
     \sul_{\substack{\alpha,\beta=1\\ \alpha<\beta}}^{N+1} (-1)^{\alpha+\beta}
          \nonumber\\
 &\hspace{2.5cm}\times
   \bigg\{
    \frac{ a(\tilde{\la}_\alpha)\, d(\tilde{\la}_\beta)}
         {\sh(\tilde{\la}_\alpha-\tilde{\la}_\beta-\eta)}
   \pl_{k=1}^{N} \big[ \sinh(\la_k-\tilde{\la}_\alpha+\eta)
                         \sinh(\la_k-\tilde{\la}_\beta-\eta)\big]
           \nonumber\\
 &\hspace{2.8cm}
   - \frac{ a(\tilde{\la}_\beta)\, d(\tilde{\la}_\alpha)}
          {\sh(\tilde{\la}_\beta-\tilde{\la}_\alpha-\eta)}
   \pl_{k=1}^{N} \big[ \sinh(\la_k-\tilde{\la}_\beta+\eta)
                         \sinh(\la_k-\tilde{\la}_\alpha-\eta)\big]
   \bigg\}
          \nonumber\\
 &\hspace{7.8cm}\times
     \det_{N-1} \Omega(\{\mu\},
          \{\tilde{\la}\}\backslash\{\tilde{\la}_\alpha,\tilde{\la}_\beta\}
                    |\{\mu\})\Bigg\}.
          \label{ffm}
\end{align}
In \eq{ffm}, we have defined again
$(\tilde{\la}_1,\ldots,\tilde{\la}_{N+1})$ to be equal to
$(\la_1,\ldots,\la_N,\xi_{m+1})$. Note that the expressions
\eq{ffp}-\eq{ffm} of the matrix elements which appear in the
summation \eq{ev-mu} are the ones that are obtained when acting on
the left with the corresponding local spin operators. This is due
to the fact that the whole algebraic procedure we have used to
derive the master equation is based from the very beginning on the
principle of action on the left.

Hence, we obtain
\be{ff-expan-s}
 \bracket{\sigma_{1,\kappa}^-\, \sigma_{\kappa, m+1}^+}
   =
  \sum_{\{\mu\}}
   \frac{\bra{\psi (\{\lambda\})}\, \sigma_{1,\kappa}^-\,
                       \ket{\psi_\kappa(\{\mu\})} \cdot
         \bra{\psi_\kappa(\{\mu\})}\, \sigma_{\kappa,m+1}^+\,
                       \ket{\psi (\{\lambda\})}}
        {\bracket{\psi (\{\lambda\})\mid\psi (\{\lambda\})}\cdot
         \bracket{\psi_\kappa(\{\mu\})\mid\psi_\kappa(\{\mu\})}
         },
\ee
where the sum is taken over all the admissible solutions
$\{\mu_1,\ldots,\mu_{N-1}\}$ of the system of $N-1$
$\kappa$-twisted Bethe equations. Observe that we did not need to
use here the completeness of the corresponding $\kappa$-twisted
Bethe states $\ket{\psi_\kappa(\{\mu\})}$ (see
\cite{TarV96,KitMST04b}) in ${\cal H}^{(M/2-N+1)}$, as the sum
over the eigenstates of ${\cal T}_\kappa$ appears automatically as
the result of the evaluation of the multiple integral \eq{master2}
by the residues outside the integration contour.

Since the resulting expression \eq{ev-mu}-\eq{ff-expan-s} is again
a polynomial in $\kappa$, we can claim from Remark~\ref{rem-pol}
that the equality \eq{ff-expan-s} is valid for all values of
$\kappa$, in particular at $\kappa= 1$. For this value of $\kappa$
and in the homogeneous limit, \eq{ff-expan-s} represents precisely
the form factor type expansion of the correlation function
$\bracket{\sigma_{1}^-\,
  \sigma_{m+1}^+}$,  with respect to all the excited
states of the Hamiltonian.

Of course, it is now clear that we can proceed in the opposite
way, starting from the form factor expansion to obtain the master
equation \eq{master2} and the re-summation of
Section~\ref{sec-31}. We will discuss this point in more details
in the next section.



\section{Dynamical master equation}\label{sec-6}

The approach described in the previous section can be generalised
to the case of time-dependent (dynamical) correlation functions.
Namely, one can derive a time-dependent analogue of the master
equations \eq{master1}-\eq{master3} for the two-point functions on
the finite chain,
\be{Defabt}
 \bracket{\sigma^\alpha_{1}(0)\,\sigma^\beta_{m+1}(t)}
   =\frac{\bra{\psi(\{\lambda\})}\,\sigma^\alpha_{1}\, e^{iHt}\,
          \sigma^\beta_{m+1}\, e^{-iHt}\,\ket{\psi(\{\lambda\})}}
         {\bracket{\psi(\{\lambda\})\mid\psi(\{\lambda\})}},
\ee
where $\ket{\psi(\{\lambda\})}$ denotes an eigenstate of the
Hamiltonian (\ref{IHamXXZ}) corresponding to an admissible
off-diagonal solution of the Bethe equations. It is possible to
obtain these time-dependent master equations either from the
formula \eq{actTM} for the multiple action of the twisted transfer
matrices or via the form factor approach. Both approaches are
described below.

\subsection{Time-dependent generating function}
\label{sec-61}

Let us first consider the dynamical two-point correlation function
of the third component of the spin in the eigenstate
$\ket{\psi(\{\lambda\})}$. Due to the property
$[\sigma_m^z,S_z]=0$, we have
\be{Defabt0}
 \bracket{\sigma^z_{1}(0)\,\sigma^z_{m+1}(t)}
=\frac{\bra{\psi(\{\lambda\})}\,\sigma^z_{1}\,e^{iH^{(0)}t}\,
       \sigma^z_{m+1}\, e^{-iH^{(0)}t}\,\ket{ \psi(\{\lambda\})}}
      {\bracket{\psi(\{\lambda\})\mid\psi(\{\lambda\})}}.
\ee
We will explain here how to derive a time-dependent generalisation
of the master equation \eq{master1} for a generating function of
\eq{Defabt0} similar to \eq{def-Q}. This dynamical generating
function is the expectation value of the following
natural time-dependent generalisation of the operator\footnote{%
In this section we consider the homogeneous chain only, since the
local Hamiltonian $H^{(0)}$ is well defined only in this case.}
\eq{GFdefQ}:
\be{Qkm} Q^\kappa_{l,m}(t)
   ={\cal T}^{l-1}\Bigl(\frac{\eta}{2}\Bigr)\cdot
    {\cal T}^{m-l+1}_\kappa\Bigl(\frac{\eta}{2}\Bigr)\cdot
    e^{ itH^{(0)}_\kappa}\cdot
    {\cal T}^{-m}\Bigl(\frac{\eta}{2}\Bigr)\cdot e^{- itH^{(0)}},
\ee
where the $\kappa$-twisted Hamiltonian $H^{(0)}_\kappa$ is defined
similarly as in \eq{ABATI}:
\be{ABATI-kap}
   H_\kappa^{(0)}=
     2\sinh\eta\,
      \frac{d{\cal T_\kappa}(\lambda)}{d\lambda}
     {\cal T}_\kappa^{-1}(\lambda)
           \pour{\lambda=\frac{\eta}{2}}
     -2M\cosh\eta.
\ee
Using the solution \eq{FCtab} of the quantum inverse problem and
the fact that the transfer matrix $\mathcal{T}_\kappa$ commutes
with the twisted Hamiltonian $H^{(0)}_\kappa$, it is easy to see
that, just like in the time-independent case, the two-point
function \eq{Defabt0} is given as
\be{cor-funct-st}
  \bracket{\sigma_{1}^z(0)\,\sigma_{m+1}^z(t)}
  =2\bracket{\sigma_{1}^z(0)}-1+
   2D^2
    \frac{\partial^2}{\partial\kappa^2} \bracket{Q^\kappa_{1,m}(t)}
    \pour{\kappa=1},
\ee
where $\bracket{Q^\kappa_{1,m}(t)}$ is the expectation value of
\eq{Qkm} in the eigenstate  $\ket{\psi(\{\lambda\})}$,
\be{def-Qt} \bracket{Q^\kappa_{1,m}(t)}=
\frac{\bra{\psi(\{\lambda\})}\,Q^\kappa_{1,m}(t)\,
\ket{\psi(\{\lambda\})}}
{\bracket{\psi(\{\lambda\})\mid\psi(\{\lambda\})}}, \ee
and $D^2$ denotes the second lattice derivative defined as in
\eq{cor-funct-ss}.

In order to be able to use Proposition \ref{lemme-actTM} to
determine the action of the operator~\eq{Qkm} on the eigenstate
$\ket{\psi(\{\lambda\})}$, one should express the twisted
evolution operator $e^{iH_\kappa^{(0)}t}$ in terms of a product of
twisted transfer matrices ${\cal T}_\kappa$. This can be done via
the Trotter type formula \cite{Tro59,Suz66},
\be{lim-U}
   e^{\pm it(H_\kappa^{(0)} +2M\cosh\eta)}=\lim_{L\to\infty}
         \left({\cal T}_\kappa\Bigl(\frac{\eta}{2}+\varepsilon\Bigr)\cdot
         {\cal T}_\kappa^{-1}\Bigl(\frac{\eta}{2}\Bigr)\right)^{\pm L},\qquad
   \varepsilon=\frac{1}{L}2it\sinh\eta.
\ee
Using now the identity
\be{T-inverse} {\cal
T}_\kappa^{-1}(\eta/2)=\Bigl(\kappa\,a(\eta/2)d(-\eta/2)\Bigr)^{-1}
{\cal T}_\kappa(-\eta/2), \ee
we arrive at the following representation for  the operator
$Q_{1,m}^\kappa(t)$:
\begin{align}
  Q_{1,m}^\kappa(t) &= \lim_{L\to\infty}
       {\cal T}_\kappa^L\Bigl(\frac{\eta}{2}+\varepsilon\Bigr)\cdot
       {\cal T}_\kappa^{m-L}\Bigl(\frac{\eta}{2}\Bigr)\cdot
       {\cal T}^{L-m}\Bigl(\frac{\eta}{2}\Bigr)\cdot
       {\cal T}^{-L}\Bigl(\frac{\eta}{2}+\varepsilon\Bigr),\\
  &= \lim_{L\to\infty}\kappa^{m-L} \,
        {\cal T}_\kappa^L\Bigl(\frac{\eta}{2}+\varepsilon\Bigr)\cdot
        {\cal T}_\kappa^{L-m}\Bigl(-\frac{\eta}{2}\Bigr)\cdot
        {\cal T}^{m-L}\Bigl(-\frac{\eta}{2}\Bigr)\cdot
        {\cal T}^{-L}\Bigl(\frac{\eta}{2}+\varepsilon\Bigr).
\end{align}
Thus, the problem of evaluating the dynamical correlation function
of the third components of the spin is reduced to the calculation
of the multiple action of twisted transfer matrices on the state
$\bra{\psi(\{\lambda\})}$. Therefore one can use directly the
results of the previous section, with the following generalisation
of Lemma~\ref{main-lem}:
\begin{lemma}\label{main-lem2}
Let $\{\lambda\}$ be an admissible off-diagonal solution of the
system of untwisted Bethe equations. There exists $\kappa_0>0$
such that, for  $0<|\kappa|<\kappa_0$, one can define a closed
contour
$\Gamma\{\lambda\}\cup\Gamma\{\eta/2\}\cup\Gamma\{-\eta/2\}$ which
satisfies the following properties:

1) it surrounds the points $\{\lambda\}$, $\eta/2$ and $-\eta/2$,
while all admissible off-diagonal solutions of the system
\eq{TTMBE_Y} are outside of this contour;

2) for $L$ large enough, the only poles which are inside and
provide non-vanishing contributions to the integral \eq{actTM} are
$z_j=\lambda_k$, $z_j=\eta/2+\varepsilon$ and $z_j=-\eta/2$;

3) for $L$ large enough, the only poles which are outside (within
a set of periodic strips) and provide non-vanishing contributions
to the integrand of \eq{actTM} are the admissible off-diagonal
solutions of the system of twisted Bethe equations \eq{TTMBE_Y}.
\end{lemma}
Proceeding then to the  limit $L\to\infty$, one obtains a
time-dependent master equation for the generating function
\eq{def-Qt}.

\begin{thm}\label{M-thm-t}
Let $\{\lambda_1,\ldots,\lambda_N\}$ be an admissible off-diagonal
solution of the system \eq{TTMBE_Y} at $\kappa=1$. Then there
exists $\kappa_0 > 0$ such that, for $0<|\kappa|< \kappa_0 $, the
generating function $\bracket{Q^\kappa_{1,m}(t)}$ \eq{def-Qt} in
the finite $XXZ$ chain \eq{IHamXXZ} is given by the multiple
contour integral
\begin{multline}\label{Master-1}
 \bracket{Q^\kappa_{1,m}(t)} =  \frac{1}{N!}
 \oint\limits_{\Gamma\{\pm\frac{\eta}2\}\cup \Gamma\{\lambda\}}
 \prod_{j=1}^{N}\frac{dz_j}{2\pi i} \cdot
 \prod_{b=1}^{N}e^{it[E(z_b)-E(\lambda_b)]
                  +im[p(z_b)-p(\lambda_b)]}
 \\
 \times
 \prod_{a=1}^{N} \frac{1}
                      { {\cal Y}_\kappa (z_a| \{z\} )}
         \cdot
 \det_N \Omega_\kappa ( \{z\} , \{\lambda\} | \{z\}  )
    \cdot
    \frac{ \det_N \Omega (\{\lambda\} , \{z\}  |\{\lambda\} )}
         { \det_N \Omega (\{\lambda\} , \{\lambda\} | \{\lambda\})}.
\end{multline}
In this expression, $E(\lambda)$ and $p(\lambda)$ denote
respectively the bare one-particle energy and momentum  \eq{E} and
\eq{P}; the integration contour is such that the only
singularities of the integrand \eq{master1} within
$\Gamma\{\pm\frac{\eta}2\}\cup \Gamma\{\lambda\}$ which contribute
to the integral  are the points $\{\pm\frac{\eta}2\}$ and
$\{\lambda\}$.
\end{thm}

Comparing the dynamical master equation \eq{Master-1} with the
corresponding result \eq{master1} in the previous section, we see
that the dependency on time appears in the integrand together with
the bare energy. In the framework of the approach described above,
this function arises as a Trotter type limit of the eigenvalues of
the operators ${\cal T}$ and ${\cal T}_\kappa$. The function
$E(z)$ has poles in the points $\pm\frac{\eta}2$, which explains
why the integration contour $\Gamma\{\pm\frac{\eta}2\}\cup
\Gamma\{\lambda\}$ in \eq{Master-1} needs to surround not only the
point $\frac\eta2$, but also the point $-\frac\eta2$.

Observe also that, unlike in the time-independent case, the
equation \eq{Master-1} holds only in a {\sl punctured} vicinity of
$\kappa=0$, i.e. for $|\kappa|$ small enough, but non-zero. This
is due to two reasons. First, the admissible solutions of the
twisted Bethe equations are separated from the point $-\eta/2$
only if $\kappa\ne 0$. Second, the expectation value
$\bracket{Q^\kappa_{1,m}(t)}$ is no longer a polynomial in
$\kappa$: due to the presence of the twisted evolution operator
$e^{itH^{(0)}_\kappa}$ in~\eq{Qkm}, it has essential singularities
at $\kappa=0,\infty$. However, it is easy to see that
$\bracket{Q^\kappa_{1,m}(t)}$ remains a holomorphic function of
$\kappa$ everywhere except at these points, which means that the
result \eq{Master-1} can be analytically continued from a
punctured vicinity of $\kappa=0$ to the whole complex plane
$\Cset^*$, and in particular to the point $\kappa=1$.

\subsection{Correlation function $\bracket{\sigma^-(0)\,\sigma^+(t)}$}
\label{sec-62}

We have seen in Section~\ref{sec-53} that the explicit expansion
of the correlation functions over the form factors can be obtained
from the master equation. It was also mentioned there that one can
follow an opposite strategy, i.e. sum up the form factors of the
local spin operators to derive the contour integrals
\eq{master1}-\eq{master3}. This way to obtain the master equation
is more direct and admits a very simple generalisation for the
time-dependent case.

Consider, for example, the dynamical correlation function
$\bracket{\sigma^-_{1}(0)\sigma^+_{m+1}(t)}$. It would seem very
natural to compute this correlation function by inserting between
the operators $\sigma^-_1(0)$ and $\sigma^+_{m+1}(t)$ the complete
set of the eigenstates of the Hamiltonian \eq{IHamXXZ}. The point
is, however, that we have no convenient parametrisation for this
complete set. Indeed, it is known that the set of the eigenstates
corresponding to the admissible off-diagonal solutions of
untwisted Bethe equations in the homogeneous case is not complete,
and that one should take into account some unadmissible solutions
as well. But, on the other hand, the vectors corresponding to
unadmissible solutions are ill-defined.

Nevertheless, we know that the  the eigenstates of the twisted
transfer matrix ${\cal T}_\kappa$ corresponding to the admissible
off-diagonal solutions of the twisted Bethe equations form a basis
of the space of states, at least for $|\kappa|$ small enough, but
non-zero. We can therefore use these states in order to sum up the
form factor series. Indeed, let us consider, similarly as in
\eq{smsp}, the $\kappa$-deformed time-dependent correlation
function
\be{kappa-g}
\bracket{\sigma^-_{1,\kappa}(0)\sigma^+_{\kappa,m+1}(t)}
  =\bracket{\sigma_{1,\kappa}^-\,e^{itH_\kappa^{(0)}-ithS_z}\,
          \sigma_{\kappa,m+1}^+\,e^{-itH^{(0)}+ithS_z}},
\ee
with
\be{kappa-lim}
\bracket{\sigma_1^-(0)\,\sigma_{m+1}^+(t)}=\lim_{\kappa\to 1}
         \bracket{\sigma^-_{1,\kappa}(0)\sigma^+_{\kappa,m+1}(t)},
\ee
where $\sigma_{1,\kappa}^-$, $\sigma_{\kappa,m+1}^+$ are defined
in \eq{deformed-spin} and $H_\kappa^{(0)}$ is given by
\eq{ABATI-kap}. Inserting now the complete set of the twisted
eigenstates between $e^{itH_\kappa^{(0)}-ithS_z}$ and
$\sigma_{\kappa,m+1}^+$ in \eq{kappa-g}, we can immediately write
the time-dependent generalisation of the equation \eq{ff-expan-s}:
\begin{multline}\label{ff-expan-st}
 \bracket{\sigma_{1,\kappa}^-(0)\, \sigma_{\kappa, m+1}^+(t)}
   =
  e^{-iht}\sum_{\{\mu\}}
 \prod_{j=1}^{N-1}e^{itE(\mu_j)} \prod_{j=1}^{N}e^{-itE(\lambda_j)}\\
   \times\frac{\bra{\psi (\{\lambda\})}\, \sigma_{1,\kappa}^-\,
                       \ket{\psi_\kappa(\{\mu\})} \cdot
         \bra{\psi_\kappa(\{\mu\})}\, \sigma_{\kappa,m+1}^+\,
                       \ket{\psi (\{\lambda\})}}
        {\bracket{\psi (\{\lambda\})\mid\psi (\{\lambda\})}\cdot
         \bracket{\psi_\kappa(\{\mu\})\mid\psi_\kappa(\{\mu\})}
},
\end{multline}
where the sum is taken over all the admissible solutions
$\{\mu_1,\ldots,\mu_{N-1}\}$ of the system of $N-1$
$\kappa$-twisted Bethe equations. It remains now to repeat all the
steps of the corresponding part of Section \ref{sec-5} in the
opposite order (from \eq{ff-expan-s} to \eq{master2}), and we
obtain
\begin{multline}\label{Master-2}
 \bracket{\sigma_{1,\kappa}^-(0)\, \sigma_{\kappa, m+1}^+(t)}=
   \frac{e^{-iht}}{(N-1)!}
\oint\limits_{\Gamma\{\pm\frac{\eta}2\}\cup \Gamma\{\lambda\}}\!
 \prod_{j=1}^{N-1}\frac{dz_j}{2\pi i} \,
 \prod_{b=1}^{N-1}e^{itE(z_b)+imp(z_b)}\prod_{b=1}^{N}
e^{-itE(\lambda_b)-imp(\lambda_b)}
\\
 \times \prod_{a=1}^{N-1} \frac{1}{ {\cal Y}_\kappa (z_a| \{z\} )}\cdot
   \pl_{j=1}^{N-1}\frac{1}{\sh(z_j-\frac\eta2)}\cdot
   \pl_{j=1}^{N}\frac{1}{\sh(\la_j-\frac\eta2)}\cdot
 \frac{\det_N \Omega ( \{\la\} , \{z\}\cup\{\frac\eta2\}\mid\{\lambda\})}
      {\det_N \Omega ( \{\la\} , \{\la\}\mid\{\lambda\})}
\\
 \times
 \Bigg\{ \sul_{\substack{\alpha,\beta=1\\ \alpha<\beta}}^{N+1}
    (-1)^{\alpha+\beta}\,
     \bigg\{
     \frac{a(\tilde{\la}_\alpha)\, d(\tilde{\la}_\beta)}
          {\sh(\tilde{\la}_\alpha-\tilde{\la}_\beta-\eta)}\cdot
   \pl_{k=1}^{N} \big[ \sinh(\la_k-\tilde{\la}_\alpha+\eta)
                         \sinh(\la_k-\tilde{\la}_\beta-\eta)\big]
\\
  \phantom{\Bigg\{ \sul_{\substack{\alpha,\beta=1\\ \alpha<\beta}}^{N+1}
           (-1)^{\alpha+\beta}\,
           \bigg\{ }
  -\frac{a(\tilde{\la}_\beta)\, d(\tilde{\la}_\alpha)}
        {\sh(\tilde{\la}_\beta-\tilde{\la}_\alpha-\eta)}\cdot
   \pl_{k=1}^{N} \big[ \sinh(\la_k-\tilde{\la}_\beta+\eta)
                         \sinh(\la_k-\tilde{\la}_\alpha-\eta)\big]\bigg\}
\\
 \times
 \det_{N-1} \Omega_\kappa ( \{z\} ,
     \{\tilde{\la}\}\backslash\{\tilde{\la}_\alpha,\tilde{\la}_\beta\}|\{z\} )
            \Bigg\}.
\end{multline}
In this expression, all the notations are just the same as in
\eq{master2}. Note that this result could have been obtained also
by the method used in Section~\ref{sec-61}.

\subsection{Dynamical correlation functions in the thermodynamic limit}
\label{sec-th}

It was explained in Remark~\ref{rem-or-series} how one could,
starting from the time-independent master equations, reproduce the
integral representations \eq{Q-int-rep}-\eq{Os+s-} for the
two-point functions in the thermodynamic limit. A similar method
can be applied to the time-dependent case \cite{KitMST04b},
although the existence of the essential singularities at
$\pm\eta/2$ in the integrand  makes this procedure more subtle.

For simplicity, we consider here only the case of the generating
function $\bracket{Q^\kappa_{1,m}(t)}$, $\ket{\psi(\{\la\})}$
being now the ground state of the Hamiltonian \eq{IHamXXZ}.
Similarly as in Remark~\ref{rem-or-series}, we can decompose the
multiple integral in \eq{Master-1} as

\be{Split-int} \oint\limits_{\Gamma\{\pm\frac{\eta}2\}\cup
\Gamma\{\lambda\}}
 \prod_{j=1}^{N}dz_j
=\sum_{n=0}^N C_N^n \oint\limits_{\Gamma\{\pm\frac{\eta}2\}}
\prod_{j=1}^{n}dz_j \oint\limits_{\Gamma\{\lambda\}}
\prod_{j=n+1}^{N}dz_j. \ee
and evaluate the integrals over the contour $\Gamma\{\lambda\}$,
which leads to a sum over the partitions of the set $\{\lambda\}$
into two disjoint subsets
$\{\lambda\}=\{\lambda_{\alpha_+}\}\cup\{\lambda_{\alpha_-}\}$. In
the time-independent case, we could then set $d(z)=0$ directly in
the remaining part of the integrand, since all the terms
proportional to $d(z)$ were holomorphic in the point $\eta/2$. In
the time-dependent case, the situation is a priori different: due
to the essential singularities of the integrand in the points
$\pm\eta/2$, the contribution of the function $d(z)$ in the
vicinity of $\eta/2$ (respectively $a(z)$ in the vicinity of
$-\eta/2$) does not vanish. Nevertheless, since $d(z)$ and $a(z)$
have zeros of order $M$ at $z=\eta/2$ and $z=-\eta/2$
respectively, one can show that the contributions of the
corresponding terms to the total result are bounded by $C^N/N!$,
where $C$ is some constant. Hence, these contributions vanish in
the thermodynamic limit $M,N\to\infty$, $M/N=\text{const}$. Thus,
in the thermodynamic limit, one can set $d(z)=0$ in the vicinity
of $z=\eta/2$ and $a(z)=0$ in the vicinity of $z=-\eta/2$. This
leads us to the following integral representation, in which the
integrand is defined differently in the two half-planes of the
complex plane \cite{KitMST04b}:
\begin{multline}\label{answer}
\bracket{Q^\kappa_{1,m}(t)}=
\sum_{n=0}^{\infty}\frac{1}{(n!)^2}\int\limits_{\cal C}d^n\lambda
\oint\limits_{\Gamma\{\pm\frac\eta2\}}\prod_{j=1}^{n}
\frac{dz_j}{2\pi i}\cdot \prod_{a,b=1}^n\frac{
\sinh(\lambda_a-z_b+\eta)\sinh(z_b-\lambda_a+\eta)}
{\sinh(\lambda_a-\lambda_b+\eta)\sinh(z_a-z_b+\eta)}
          \\
\times
\prod_{b=1}^ne^{it[E(z_b)-E(\lambda_b)]+im[p(z_b)-p(\lambda_b)]}\cdot
\det_n M_{\kappa}(\{\lambda\},\{z\}) \cdot\det_n[{\cal
R}^{\kappa}_n(\lambda_j,z_k|\{\lambda\},\{z\})].
\end{multline}
In this expression, the contour $\Gamma\{\pm\frac\eta2\}$
surrounds the points $\pm\frac\eta2$ and does not contain any
other singularities of the integrand.

Comparing this result with its time-independent analogue
\eq{Q-int-rep}, we see that the determinant of densities
$\det[\rho(\la_j,z_k)]$ has been replaced by a new thermodynamic
quantity depending on the function ${\cal R}^{\kappa}_n(\lambda,z|%
\{\lambda_1,\dots,\lambda_n\},\{z_1,\dots,z_n\})$. This function
is defined differently in the vicinities of $\eta/2$ and
$-\eta/2$:
\be{cal-R} {\cal R}^{\kappa}_n(\lambda,z|
\{\lambda\},\{z\})=\left\{
\begin{array}{l}
\rho(\lambda,z),\qquad z\sim\eta/2;\\
-\kappa^{-1}\rho(\lambda,z+\eta)
\prod\limits_{b=1}^{n}\frac{\sinh(z-\lambda_b+\eta)
\sinh(z_b-z+\eta)}{\sinh(\lambda_b-z+\eta)\sinh(z-z_b+\eta)},
\qquad z\sim-\eta/2.
\end{array}\right.
\ee
Due to the factors $\exp(itE(z_b))$, the integrand in \eq{answer}
has  essential singularities in the points $\pm\eta/2$. However,
in the case $t=0$, these essential singularities disappear and the
integrals around $-\eta/2$ vanish. The remaining part of the
integrand has poles of  order $m$ at $z_j=\eta/2$. Hence, at
$t=0$, the sum over $n$ in \eq{answer} is actually restricted to
$n\le m$, and we reproduce the result of \eq{Q-int-rep}.

\section*{Conclusion}

In this review, we have summarised recent results concerning the
computation of correlation functions in the $XXZ$ chain by the
methods of the inverse scattering problem and the algebraic Bethe
Ansatz. In conclusion, we would like to discuss some perspectives
and problems to be solved.

One of the most interesting open problems is to prove the
conformal field theory predictions concerning the asymptotic
behaviour of the correlation functions. We have seen in
Section~\ref{sec-4} that the new integral representations
presented in Section~\ref{sec-3} seemed, from the view point of
the asymptotic analysis, more promising than the original
representations in terms of the elementary blocks of
Section~\ref{sec-2}. Nevertheless, it is still not clear how to
extract the asymptotics directly from these multiple integrals.

A possible way to solve this problem would be to find the
thermodynamic limit of the master equations. It is natural to
expect that, in this limit, one should obtain a representation for
the two-point functions in terms of a single functional integral,
which could probably  be estimated for the large time and
distance.

Irrespective of this problem of the asymptotics, the master
equation shows a direct analytic relation between the multiple
integral representations and the form factor expansions for the
correlation functions. It seems likely that similar
representations exist for other model solvable by  algebraic Bethe
Ansatz. It would be in particular very interesting to obtain an
analogue of this master equation in the case of the field theory
models, which could provide an analytic link between short
distance and long distance expansions of the correlation
functions.

Another interesting further development would be to generalise all
these results to the case of correlation functions at finite
temperature. In this direction, a multiple integral representation
similar to \eq{Q-int-rep} was derived recently in \cite{GohKS04}
for the temperature time-independent correlation function, and it
would be interested to obtain it in the time-dependent case as
well. One can also wonder whether there exists a master equation
similar to \eq{master1} for the temperature-dependent case. It
would raise the interesting question of the form factor expansion
at non-zero temperature.

It is also well known that, for the case of free fermions
$\Delta=0$, the dynamical correlation functions of the $XXZ$ chain
satisfy difference-differential classical exactly solvable
equations \cite{MccTW77,Per80,ItsIKS93}. It is natural to wonder
whether this property holds also for general $\Delta$, or at least
for some specific cases. We hope that the multiple integral
representations for the dynamical correlation functions open a way
to study this problem.

\section*{Acknowledgements}
J. M. M., N. S. and V. T. are supported by CNRS.
N. K., J. M. M. and V. T. are supported by the European network
EUCLID-HPRNC-CT-2002-00325.
J. M. M. and N. S. are supported by INTAS-03-51-3350.
N. S. is supported by the French-Russian Exchange Program,
the Program of RAS Mathematical Methods of the Nonlinear Dynamics,
RFBR-02-01-00484, Scientific Schools 2052.2003.1.
N. K, N. S. and V. T. would like to thank the Theoretical Physics group
of the Laboratory of Physics at ENS Lyon for hospitality, which makes
this collaboration possible.
All the authors would like to thank the organisers of the RIMS COE research
program in Kyoto for their invitation.


\bibliographystyle{../../tex/TeX/style-biblio/h-physrev}

\bibliography{../../tex/bibliographie/biblio}

\end{document}